\documentclass[sigconf]{acmart}
\AtBeginDocument{%
  }

\usepackage{eso-pic}

\AddToShipoutPictureBG*{%
  \AtPageUpperLeft{%
    \put(0,-30){%
      \parbox{\paperwidth}{\centering\footnotesize\bfseries Accepted by The Web Conference (WWW) 2026}%
    }%
  }%
}

\copyrightyear{2026}
\acmYear{2026}
\setcopyright{cc}
\setcctype{by}
\acmConference[WWW '26] {Proceedings of the ACM Web Conference 2026}{April 13--17, 2026}{Dubai, United Arab Emirates.}
\acmBooktitle{Proceedings of the ACM Web Conference 2026 (WWW '26), April 13--17, 2026, Dubai, United Arab Emirates}
\acmISBN{979-8-4007-2307-0/2026/04}
\acmDOI{10.1145/XXXXXX.XXXXXX}

\usepackage{multicol}
\usepackage{xspace}
\usepackage{enumerate}
\usepackage{adjustbox}
\usepackage{multirow}
\usepackage{subfigure}
\usepackage{color}
\usepackage{soul}
\usepackage{makecell}
\definecolor{MyRed}{RGB}{151,0,0}
\definecolor{Yellow}{RGB}{243,140,10}
\usepackage{booktabs} 
\usepackage{enumerate}
\usepackage{booktabs}
\usepackage{makecell}
\usepackage{multirow}
\usepackage{flushend}
\usepackage{latexsym}
\usepackage{multirow}
\usepackage{amsmath}
\usepackage{amsfonts}
\usepackage[scr=dutchcal,  
            cal=cm]   
           {mathalpha}
\usepackage{caption}
\usepackage{subcaption}
\usepackage[noend]{algpseudocode}
\usepackage{graphicx}
\usepackage{color}
\usepackage{algpseudocode}  
\usepackage[linesnumbered, ruled, vlined, shortend]{algorithm2e}
\usepackage{verbatim}
\usepackage[graphicx]{realboxes}
\usepackage{rotating}
\usepackage{array}
\usepackage{threeparttable}
\usepackage{soul}
\usepackage{tabularx}
\usepackage{xcolor}
\usepackage[most]{tcolorbox}
\tcbuselibrary{breakable}
\usepackage{caption}
\usepackage{enumitem}
\usepackage[table]{xcolor}
\usepackage{balance}

\newcommand{\framework}{RagSEDE}

\begin{document}

\title{Effective and Unsupervised Social Event Detection and Evolution via RAG and Structural Entropy}


\author{Qitong Liu}
\orcid{0009-0008-7135-5104}
\affiliation{%
  \institution{Beihang University}
  \state{Beijing}
  \country{China}
}
\email{liuqt@buaa.edu.cn}

\author{Hao Peng}
\authornote{Corresponding author.}
\orcid{0000-0003-0458-5977}
\affiliation{%
  \institution{Beihang University}
  \state{Beijing}
  \country{China}
}
\email{penghao@buaa.edu.cn}

\author{Zuchen Li}
\orcid{0009-0000-3819-833X}
\affiliation{%
  \institution{Beihang University}
  \state{Beijing}
  \country{China}
}
\email{lizuchen@buaa.edu.cn}

\author{Xihang Meng}
\orcid{0009-0000-4545-1620}
\affiliation{%
  \institution{Beihang University}
  \state{Beijing}
  \country{China}
}
\email{xihangmeng@buaa.edu.cn}

\author{Ziyu Yang}
\orcid{0009-0004-3049-1768}
\affiliation{%
  \institution{Beihang University}
  \state{Beijing}
  \country{China}
}
\email{yangziyu@buaa.edu.cn}

\author{Jiting Li}
\orcid{0000-0002-2625-0935}
\affiliation{%
  \institution{Academy of Military Sciences}
  \state{Beijing}
  \country{China}
}
\email{lijiting_1993@126.com}

\author{Li Sun}
\orcid{0000-0003-4562-2279}
\affiliation{%
  \institution{North China Electric Power University}
  \state{Beijing}
  \country{China}
}
\email{ccesunli@ncepu.edu.cn}

\author{Philip S. Yu}
\orcid{0000-0002-3491-5968}
\affiliation{%
  \institution{University of Illinois Chicago}
  \state{Chicago}
  \country{USA}
}
\email{psyu@uic.edu}

\settopmatter{authorsperrow=4}

\renewcommand{\shortauthors}{Qitong Liu et al.}

\begin{abstract}
  With the growing scale of social media, social event detection and evolution modeling have attracted increasing attention. 
Graph neural networks (GNNs) and transformer-based pre-trained language models (PLMs) have become mainstream approaches in this area. 
However, existing methods still face three major challenges. 
First, the sheer volume of social media messages makes learning resource-intensive.
Second, the fragmentation of social media messages often impedes the model's ability to capture a comprehensive view of the events. 
Third, the lack of structured temporal context has hindered the development of effective models for event evolution, limiting users' access to event information.
To address these challenges, we propose a foundation model for unsupervised \underline{S}ocial \underline{E}vent \underline{D}etection and \underline{E}volution, namely \framework. 
Specifically, \framework~ introduces a representativeness- and diversity-driven sampling strategy to extract key messages from massive social streams, significantly reducing noise and computational overhead.
It further establishes a novel paradigm based on Retrieval Augmented Generation (RAG) that enhances PLMs in detecting events while simultaneously constructing and maintaining an evolving event knowledge base. 
Finally, \framework~ leverages structural information theory to dynamically model event evolution keywords for the first time.
Extensive experiments on two public datasets demonstrate the superiority of \framework~ in open-world social event detection and evolution. 
\end{abstract}




\maketitle

\section{Introduction}
\label{sec:introduction}
Social media platforms have become a vital source of real-time signals for emerging events, drawing increasing attention to the task of event analysis from large-scale, unstructured social message streams~\cite{qian2023open, xian2025community}. 
Social Event Detection (SED) focuses on automatically identifying and clustering social messages that pertain to the same real-world event~\cite{cao2021knowledge}. 
Complementing this, Social Event Evolution (SEE) aims to characterize the temporal dynamics of an event, capturing how its semantic content evolves over time~\cite{peng2021streaming}.
Together, SED and SEE support a wide range of downstream applications, including community discovery~\cite{fortunato202220}, recommender systems~\cite{zhao2024recommender, rajput2023recommender}, and information retrieval~\cite{qian2022integrating, dai2024bias}.

\begin{figure}[h]
    \centering
    \includegraphics[width=1\linewidth]{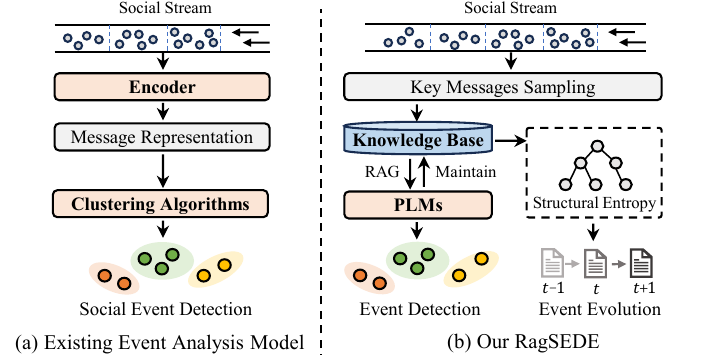}
    \caption{Illustration comparison of the existing event analysis model and our proposed~\framework{}.}
    \label{fic: introduction}
\end{figure}

As shown in Figure~\ref{fic: introduction}(a), recent models for social event analysis primarily focus on event detection. 
These models typically follow a representation–clustering paradigm, where social messages are first encoded into latent representations and then clustered, with each cluster corresponding to a distinct event. 
Several studies~\cite{peng2019fine, cui2021mvgan, ren2022known, cao2021knowledge, ma2025enhanced} leverage heterogeneous interactions (e.g., users, entities, hashtags) to model the structural dependencies among social messages. 
Furthermore, other works~\cite{yu2025promptsed, li2024relational} utilize the powerful contextual semantics provided by PLMs such as BERT~\cite{devlin2019bert} and SBERT~\cite{reimers2019sentence} to encode social messages. 
Most of these methods~\cite{peng2022reinforced, cao2021knowledge} use K-Means or DBSCAN~\cite{ester1996density} to cluster message representations, and~\cite{cao2024hierarchical} employs a structural entropy minimization algorithm~\cite{li2016structural} to achieve unsupervised clustering.

However, existing approaches face several critical challenges in the context of open-world, unsupervised social event modeling. 
\textbf{First}, the massive volume and informal nature of social messages lead to resource-intensive computation and high noise levels. 
For popular events, there are often large numbers of semantically similar or even identical messages.
\cite{ren2022known, cao2021knowledge, cao2024hierarchical} use all messages for event detection, resulting in significant resource waste. 
Furthermore, users often post messages randomly, and many messages lack attributes, leading to missing or noisy edges in GNN-based methods.
Although~\cite{yu2025towards} employs message anchors to reduce computational overhead, it still relies on building a global graph using all messages with noisy edges.
\textbf{Second}, the fragmented nature of social messages often prevents models from capturing a complete picture of events, resulting in shallow representations. 
As shown in Figure~\ref{fic: introduction}(a), existing methods rely solely on message-level embeddings during the clustering process, without incorporating any global semantic guidance from previously occurred events. 
Ignoring these global signals hinders the model’s ability to align new messages with the broader narrative, particularly in cases involving ambiguous phrasing or topic drift~\cite{knights2009detecting}.
\textbf{Third}, most current works focus only on event detection, overlooking the temporal dynamics and evolution of events, which are crucial for understanding how narratives unfold over time. 
Although~\cite{liu2020event} considers the evolution of sub-event structure over time, it ignores the evolution of event semantics.
This limitation significantly restricts users’ access to time-sensitive event knowledge.

To tackle these three aforementioned challenges, we propose a novel foundation model for unsupervised \underline{S}ocial \underline{E}vent \underline{D}etection and \underline{E}volution, namely \framework. 
Our proposed framework, as illustrated in Figure~\ref{fic: introduction}(b), consists of three parts: key message sampling, RAG-based event detection, and structural entropy-based event evolution.
\textbf{First}, to mitigate the computational inefficiency caused by highly semantically similar messages, \framework{} employs a sampling strategy that balances representativeness and diversity to extract key messages from the social message stream. 
Using only these key messages rather than the entire message set for event detection, \framework{} achieves efficient detection in open-world settings.
\textbf{Second}, leveraging the powerful contextual reasoning capabilities of PLMs, \framework{} introduces a novel RAG-based event detection paradigm. 
Unlike prior methods that follow a "representation-then-cluster" approach, \framework{} constructs and maintains a structured knowledge base of detected events and employs RAG~\cite{lewis2020retrieval,shi2024generate,fan2024survey,zhao2024retrieval} to inject global semantic signals into PLMs during detection. 
Furthermore, because~\framework{} relies solely on the semantic content of messages, it avoids the issues of missing or noisy edges mentioned above.
\textbf{Third}, by integrating the constructed knowledge base with structural information theory~\cite{li2016structural}, \framework{} dynamically model the semantic evolution of events over time. 
Structural information theory~\cite{li2016structural} suggests that minimizing structural entropy can reveal the essential information embedded in the graph. 
Based on this principle, \framework{} introduces graph construction, inheritance, and forgetting mechanisms, enabling it to extract temporally evolving event keywords by minimizing the structural entropy.
The codes of ~\framework{} are publicly available on GitHub\footnote{https://github.com/SELGroup/RagSEDE}.
In summary, the contributions of this paper are as follows:

$\bullet$
We propose a novel framework for streaming, unsupervised social event detection and evolution, termed~\framework, which integrates RAG and structural information theory to effectively address three key challenges in open-world social event analysis.

$\bullet$
We propose a novel paradigm for event detection that leverages knowledge bases and RAG, fully exploiting the powerful contextual understanding capabilities of PLMs. 
By incorporating a key message sampling strategy, our method significantly reduces computational costs while maintaining detection effectiveness.

$\bullet$
We model the dynamic evolution of event keywords.
By minimizing structural entropy over time, our framework effectively captures the evolving semantic cores of events.

$\bullet$
Extensive experiments on two benchmark datasets demonstrate that~\framework{} consistently outperforms strong baselines in both detection accuracy and evolution summarization.

\section{Preliminaries}
\label{sec:problem_definition}
This section summarizes some important concepts and definitions.
The glossary of notations and the detailed introduction of structural entropy are provided in Appendix~\ref{AppdxGN} and Appendix~\ref{AppdxSE}, respectively.

\subsection{SED}
\label{subsec:SED}
A social stream $S$ is defined as a continuous and temporal sequence of message blocks, that is, $S = \{\text{M}_1,\ \text{M}_2,\ \cdots\}$, where block $\text{M}_t = \{m_i \mid 1 \leq i \leq |\text{M}_t|\}$ contains all messages that arrive within the time interval $[t,\ t+1)$.
The SED task is to extract clusters of correlated messages from $S$ to represent real-world events.
Finally, each $m_i$ corresponds to an event label $y_{m_i}$ with which it is associated.

\subsection{Structural Entropy}
\label{subsec:SE}
Structural entropy~\cite{li2016structural} is a measure of the uncertainty of a graph structure. 
It is computed based on the graph's 2D encoding tree, where non-leaf nodes represent graph partitions.
Given a graph $G$ and its encoding tree $\mathcal{T}$, the 2D structural entropy is defined as:
\begin{equation}
H^\mathcal{T}(G) = -\sum_{\alpha \in \mathcal{T},\ \alpha \ne \lambda}\frac{g_\alpha}{vol(G)}\log_2\frac{vol(\alpha)}{vol(\alpha^-)},
\end{equation}
where $\alpha$ is a non-root node of $\mathcal{T}$, $g_\alpha$ and $vol(\alpha)$ is the cut degree and volume of $\alpha$, and $\alpha^-$ is the parent node of $\alpha$.
\cite{li2016structural} proposes a vanilla greedy 2D structural entropy minimization algorithm, which repeatedly merges any two nodes in the encoding tree until structural entropy reaches the minimum possible value.

\section{Methodology}
\label{sec:methodology}
\begin{figure*}[t]
\centering
\includegraphics[width=1\linewidth,trim={0cm 0 0cm 0}, clip]{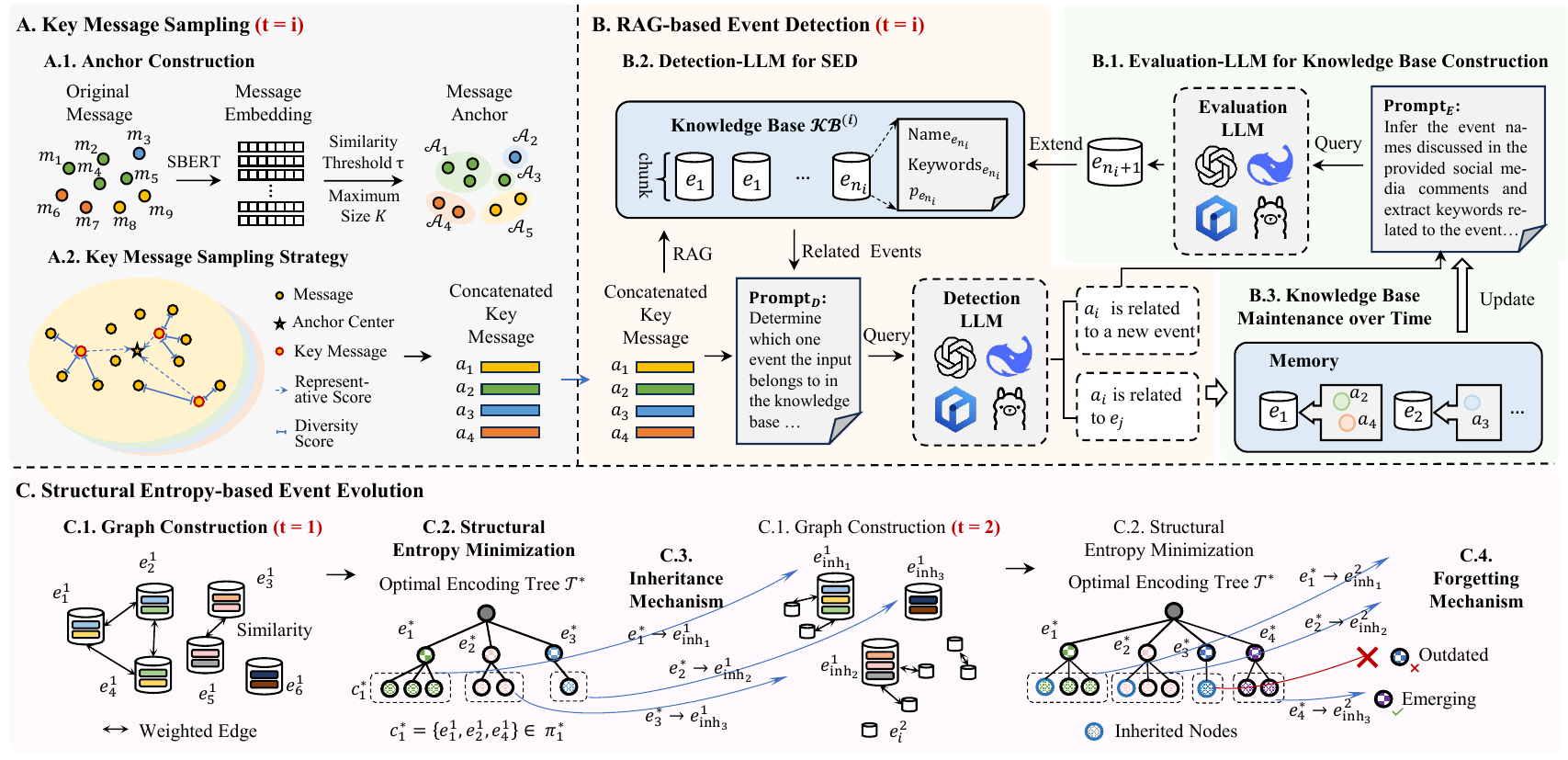}
\caption{The proposed \framework~framework.
}
\label{fig:model}
\end{figure*}

In this section, we systematically describe the proposed~\framework{}.
Section~\ref{subsec:KMS} presents a strategy for sample key messages. Section~\ref{subsec:SED} introduces the event detection paradigm of ~\framework{} based on RAG-enhanced PLMs. Section~\ref{subsec:SEE} introduces the event evolution method based on structural entropy minimization.

\subsection{Key Message Sampling (KMS)}
\label{subsec:KMS}
Given a message block M$_t$ in a high-throughput social stream, we select a set of key messages from M$_t$ that preserves both the representativeness and the diversity within each anchor, as shown in Figure~\ref{fig:model}(A). 
This approach reduces computation while maintaining detection quality for downstream detection and evolution. 

\subsubsection{Anchor Construction}
\label{subsubsec:AC}
This step aims to aggregate messages that are highly semantically similar or even exactly identical into anchors.
We first encode each message $m_i$ into a $z_i$.
Specifically,
\begin{equation}
    z_i = \mathrm{Enc}(m_i) \in \mathbb{R}^d,
\end{equation}
where $\mathrm{Enc}(\cdot)$ denotes a sentence encoder, which we use SBERT here, and $d$ is the hidden state dimension of SBERT.
Then, we compute the cosine similarity between two messages, specifically,
\begin{equation}
    s_{ij} \;=\; \frac{z_i^\top z_j}
    {\|z_i\|_2 \, \|z_j\|_2},
\end{equation}
where $\| \cdot \|$ represents the euclidean norm.
Messages are assigned to the same anchor \(\mathcal{A}_k\) if and only if
\begin{equation}
    s_{ij} \ge \tau, \quad \forall m_i, m_j \in \mathcal{A}_k,
\end{equation}
where \(\tau \in (0,1)\) is a predefined similarity threshold.
To avoid oversized anchors for popular events, we set a maximum size $K$ to limit the number of messages in each anchor.
If adding a new message would cause $|\mathcal{A}_k| > K$, a new anchor is created.
All messages can thus be represented as an anchor collection $\mathbb{A} = \{\mathcal{A}_1, \mathcal{A}_2, \dots \}$.

\subsubsection{Key Message Sampling Strategy}
\label{subsubsec:KMSS}
To select key messages from each anchor, we define two scoring criteria.
\paragraph{Representativeness Score.}
The representativeness score measures how well a message reflects the
central semantics of its anchor. 
Formally, for a $m_i$ in $\mathcal{A}_k$, representativeness score is defined as
\begin{equation}
    \mathrm{Rep}(m_i) = \frac{z_i^\top c_k}{\|z_i\|_2 \, \|c_k\|_2}, \quad c_k = \frac{1}{n} \sum _{j=1} ^{n} z_j,
\end{equation}
where $z_i$ is the embedding of $m_i$, $c_k$ is the center embedding of $\mathcal{A}_k$, and $n$ is the message number of $\mathcal{A}_k$.
A high representativeness score
indicates strong semantic alignment with the anchor center, thereby capturing the dominant event of the anchor.
\paragraph{Diversity Score.}
The diversity score measures the degree to which a message provides novel information from the same anchor.
For a message $m_i$ in anchor $\mathcal{A}_k$, its diversity score is defined as
\begin{equation}
    \mathrm{Div}(m_i) = 
    \frac{1}{n-1} \sum_{m_j \in \mathcal{A}_k, j \ne i}
    \frac{z_i^\top z_j}
    {\|z_i\|_2 \, \|z_j\|_2},
\end{equation}
where $z_i$ is the embedding of $m_i$, and $n$ is the message number of $\mathcal{A}_k$.
A high diversity score means different aspects of the event.
\paragraph{Combined Score.}
The final score balances the two criteria:
\begin{equation}
    \mathrm{S}(m_i) = 
    \lambda \cdot \mathrm{Rep}(m_i) + 
    (1-\lambda) \cdot \mathrm{Div}(m_i),
\end{equation}
where $\lambda \in [0,1]$ controls the trade-off.
At each anchor $\mathcal{A}_k$, we select the top $p$ messages 
with the highest combined scores as key messages and concatenate them as $a_k$ for SED and SEE.

\subsection{RAG-based Event Detection}
\label{subsec:SED}
In this section, we present our RAG-based event detection framework, as shown in Figure~\ref{fig:model}(B), which is designed to incorporate both the construction and maintenance of an event knowledge base (Sections~\ref{subsubsec:ELLM} and~\ref{subsubsec:KBM}) and the use of RAG to leverage the knowledge base as external global guidance for event detection (Section~\ref{subsubsec:DLLM}).
To ensure computational efficiency, we rely not on all messages but only on the key messages sampled in Section~\ref{subsec:KMS}.
Take M$_t$ as an example, the final input set for detection is thus $A = \{a_1, a_2, \dots \}$, where each $a_k$ serves as the detection unit in subsequent steps.
We present the algorithm for this section in Algorithm~\ref{alg:rag-sed}. 

\subsubsection{Evaluation-LLM for Knowledge Base Construction}
\label{subsubsec:ELLM}
We dynamically build and expand a knowledge base during detection. 
This knowledge base stores structured information about previously detected events, including their names, salient keywords, and embeddings. 
Formally, the knowledge base for M$_t$ is defined as:
\begin{equation}
    \mathcal{KB}^{(t)} = \{e = (\mathrm{Name}_e,\ \mathrm{Keywords}_e,\ p_e)\},
\end{equation}
\begin{equation}
    p_e = \mathrm{Enc}(\mathrm{Name}_e\ ||\ \mathrm{Keywords}_e) \in \mathbb{R}^d,
\end{equation},
where $\mathrm{Name}_e$ denotes the name of $e$, $\mathrm{Keywords}_e = \{w_1,\ w_2,\\ \dots\}$ is the keyword set of $e$, and $p_e \in \mathbb{R}^d$ is the event embedding. 
The embedding is obtained by encoding both the name and keywords.
Each $e$ is stored as an independent chunk in the knowledge base, enabling efficient retrieval and expansion during detection.

The event information, i.e., $\mathrm{Name}_e$ and $\mathrm{Keywords}_e$, is extracted from key messages using an Evaluation-LLM guided by our designed $\mathrm{Prompt}_E$.
The $\mathrm{Prompt}_E$ includes (i) a task description (``infer the event names discussed in the provided social media comments and extract keywords related to the event''), (ii) specific execution rules, and (iii) output requirements specifying a structured JSON format.
We provide the complete $\mathrm{Prompt}_E$ in Appendix~\ref{AppdxPromptE}.

When new messages $a_k$ arrive and cannot be aligned with any existing event in the knowledge base, the Evaluation-LLM is again invoked to generate a new event chunk:
\begin{equation}
\label{eq:ELLM3}
    e_{\text{new}} = \mathrm{LLM}_{\mathrm{Prompt}_E}(a_k)
    = (\mathrm{Name}_{\text{new}},\ \mathrm{Keywords}_{\text{new}},\ p_{\text{new}}),
\end{equation}
where $a_k$ is the aggregated message from the new anchor.
The new event $e_{\text{new}}$ is then inserted into the knowledge base:
\begin{equation}
\label{eq:ELLM4}
    \mathcal{KB}^{(t)} \;\leftarrow\; \mathcal{KB}^{(t)} \cup \{ e_{\text{new}} \}.
\end{equation}
Through this incremental process, the knowledge base gradually accumulates structured information about various events, providing powerful global semantic guidance for detection.

\begin{algorithm}[t]
\caption{RAG-based Event Detection of M$_t$}
\label{alg:rag-sed}
\KwIn{Key message set $A = \{a_1, a_2, \dots\}$, empty knowledge base $\mathcal{KB}^{(t)}$, buffer threshold $\theta$}
\KwOut{Event label $\{y_{m_i}\}$ for each $m_i$ in M$_t$}

\For{each aggregated message $a_k \in A$}{
    \tcp{--- Step 1: Retrieval ---}
    Encode $a_k$ and compute similarity $r(e \mid a_k)$ for all $e \in \mathcal{KB}^{(t)}$ via Eq.\ref{eq:DLLM1}\;
    Select top-$q$ events above threshold via Eq.\ref{eq:DLLM2}\;
    
    \tcp{--- Step 2: Detection ---}
    Query Detection-LLM via Eq.\ref{eq:DLLM3}\;
    \eIf{Detection-LLM outputs $e \in \mathcal{N}_{E(a_k)}$}{
        Assign $y_{a_k} = e$ and append $a_k$ to buffer $\mathcal{B}_e$\;
    }{
        \tcp{New event detected}
        $e_{\text{new}} \gets$ Query Evaluation-LLM via Eq.\ref{eq:ELLM3}\;
        Update $\mathcal{KB}^{(t)}$ via Eq.\ref{eq:ELLM4}\;
        Initialize buffer $\mathcal{B}_{e_{\text{new}}} \gets \{a_k\}$\;
        Assign $y_{a_k} = e_{\text{new}}$\;
    }
    
    \tcp{--- Step 3: Knowledge Base Maintenance ---}
    \For{each $e \in \mathcal{KB}^{(t)}$}{
        \If{$|\mathcal{B}_e| \ge \theta$}{
            Generate refreshed $(\mathrm{Keywords}_e,\ p_e)$ via $\mathrm{LLM}_{\mathrm{Prompt}_E}(\mathcal{B}_e)$ in Eq.\ref{eq:ELLM3}\;
            Update $\mathcal{KB}^{(t)}$ via Eq.\ref{eq:ELLM4}\;
            Clear buffer $\mathcal{B}_e$\;
        }
    }
}
\Return{$\{y_{m_i}\} \gets$ Eq.\ref{eq:DLLM4}}
\end{algorithm}

\subsubsection{Detection-LLM for SED}
\label{subsubsec:DLLM}
To enhance the capability of LLMs in social event detection, we design a RAG-based paradigm.
Given an incoming aggregated message $a_k$, RAG first queries the knowledge base $\mathcal{KB}^{(t)}$ to retrieve the most semantically relevant candidate events by event embedding.
Retrieval of RAG is performed using cosine similarity.
Formally, for each event $e$ in $\mathcal{KB}^{(t)}$:
\begin{equation}
\label{eq:DLLM1}
r(e \mid a_k) =
\frac{\mathrm{Enc}(a_k)^\top p_e}{\| \mathrm{Enc}(a_k)\|_2 \cdot \|p_e\|_2},
\end{equation}
where $r(e \mid a_k)$ represents the degree of correlation between event $e$ and message $a_k$ in the knowledge base, $p_e$ represents the embedding of $e$, and $\mathrm{Enc}$ is the embedding model used to obtain $p_e$.
Sort the correlation of all events in the knowledge base from high to low, and the top-$q$ events are selected as the retrieval events:
\begin{equation}
\label{eq:DLLM2}
\mathcal{N}_{E(a_k)} = 
\underset{e \in \mathcal{KB}^{(t)}}{\operatorname{arg\,max}^{\,q}} \; 
\{ r(e \mid a_k) \;|\; r(e \mid a_k) \ge \gamma \},
\end{equation}
where $\mathcal{N}_{E(a_k)}$ represents the retrieval event set, and $\gamma$ represents the similarity threshold of RAG.

The Detection LLM is then prompted with both the query message $a_k$ and the $\mathcal{N}_{E(a_k)}$.
Our designed $\mathrm{Prompt}_D$ consists of (i) a task description (``determine which
one event the input belongs to in the knowledge base''), (ii) specific execution
rules, (iii) output requirements specifying a structured JSON
format, and (iv) the retrieved event set as global semantic guidance. We provide the complete $\mathrm{Prompt}_D$ in Appendix~\ref{AppdxPromptD}.
The detection decision is formalized as:
\begin{equation}
\label{eq:DLLM3}
y_{a_k} = \mathrm{LLM}_{\mathrm{Prompt}_D}(a_k,\ \mathcal{N}_{E(a_k)}) = \begin{cases} 
            e \in \mathcal{N}_{E(a_k)} \\
	      \mathrm{Others}	
		\end{cases},
\end{equation}
where $y_{a_k}$ is the event $a_k$ discussed obtained by Detection-LLM based on $\mathrm{Prompt}_D$ and the knowledge base.
It is worth noting that when the retrieval event set is empty or irrelevant, the Detection-LLM output is "Others".
For any $a_k$ predicted as "Others", we treat it as belonging to a previously unseen event $e_{\mathrm{new}}$, i.e., $y_{a_k} = e_{\mathrm{new}}$.
In this case, $a_k$ is forwarded to the Evaluation-LLM (Section~\ref{subsubsec:ELLM}), which generates the event name and keywords of $e_{\mathrm{new}}$, and the knowledge base is updated accordingly to enable Detection-LLM to identify subsequent messages related to $e_{\mathrm{new}}$.

Finally, according to $y_{a_k}$, the event label of the original message is obtained as:
\begin{equation}
\label{eq:DLLM4}
y_{m_i} = y_{a_k},\quad \forall \, m_i \in \mathcal{A}_k,
\end{equation}
where $\mathcal{A}_k$ is the corresponding anchor of $a_k$.
Through this RAG-based paradigm, our framework effectively incorporates structured event information as global semantic guidance, significantly enhancing LLMs to perform accurate and efficient SED.

\subsubsection{Knowledge Base Maintenance over Time (KBM)}
\label{subsubsec:KBM}
To prevent the event information in the knowledge base from becoming outdated during long-term detection, we design a maintenance mechanism that periodically refreshes event information with new messages.
Specifically, for each event $e \in \mathcal{KB}^{(t)}$, we maintain a buffer $\mathcal{B}_e$ to record the set of messages that have been assigned to $e$ by the detection process.
When the number of messages in the buffer exceeds a predefined threshold $\theta$, the buffered messages are concatenated and forwarded to the Evaluation-LLM.
The LLM generates refreshed event keywords, and we then update $\mathcal{KB}^{(t)}$.
After the update, the buffer is cleared, and the system continues to collect messages.
Once the buffer again reaches $\theta$, the update process is repeated.
Through this threshold-triggered refresh mechanism, our framework ensures that event information remains up-to-date and robust even in high-volume, long-duration social streams.

\begin{algorithm}[t]
\caption{Structural Entropy-based Event Evolution}
\label{alg:se-see}
\KwIn{Daily knowledge bases $\{\mathcal{KB}^{(1)}, \dots, \mathcal{KB}^{(T)}\}$}
\KwOut{Aligned events $\{\mathcal{E}^{*(1)}, \dots, \mathcal{E}^{*(T)}\}$}

\For{$t \gets 1$ \KwTo $T$}{
    \tcp{--- Step 1: Graph Construction ---}
    Construct graph $G_t$ with $\mathcal{E}_\mathrm{inh}^{t-1}$ via Eq.\ref{eq:SEE1} and Eq.\ref{eq:SEE2}\;

    \tcp{--- Step 2: Structural Entropy Minimization ---}
    Enforce no-merge constraint for inherited nodes\;
    Apply structural entropy minimization in $G_t$\;
    Obtain aligned events $\mathcal{E}^{*(t)} = \{e_1^*, e_2^*, \dots\}$\;

    \tcp{--- Step 3: Inheritance Mechanism ---}
    $\mathcal{E}_\mathrm{inh}^{t} \gets \mathcal{E}^{*(t)}$\;

    \tcp{--- Step 4: Forgetting Mechanism ---}
    \For{each $e_i^* \in \mathcal{E}^{*(t)}$}{
          \If{$e_i^*$ is not supported by ordinary nodes in partitioning}{
          $\mathcal{E}_\mathrm{inh}^{t} \gets\mathcal{E}_\mathrm{inh}^{t} \setminus e_i^*$\;
          }
        }
}
\Return{$\{\mathcal{E}^{*(1)}, \dots, \mathcal{E}^{*(T)}\}$}
\end{algorithm}

\subsection{Structural Entropy-based Event Evolution}
\label{subsec:SEE}
In this section, we present a structural entropy-based SEE framework that tracks changes of event keywords over time, as shown in Figure~\ref{fig:model}(C).
After daily SED, we obtain a structured knowledge base containing event names and their keywords.
However, these knowledge bases cannot directly capture event evolution due to two issues: 
(i) Misaligned event granularity. 
During detection, for popular events, the LLM tends to generate fine-grained subcategories; conversely, for cold events, it often generates coarse-grained categories. 
(ii) Daily initialization. Since event detection is performed with a newly initialized knowledge base each day, it becomes difficult to determine which events in different daily knowledge bases correspond to the evolution of the same real-world event.
To address these challenges, we design an SEE framework comprising four components: graph construction, structural entropy minimization, an inheritance mechanism, and a forgetting mechanism. 
The first two components align event granularity, while the latter two ensure temporal continuity of evolving events. 
We present the algorithm for this section in Algorithm~\ref{alg:se-see} and describe it as follows.

\begin{table*}[t]
  \centering
  \footnotesize
  \setlength{\tabcolsep}{4.77pt}
  \renewcommand{\arraystretch}{0.7}
  \caption{Social event detection results on Events2012. * marks results acquired with the ground truth message labels. The best results are \textbf{bolded}, the second-best results are \underline{underlined}, and the proposed method is marked in an orange background.}
  \label{table:openset_results_event2012}
    \begin{threeparttable}
    \begin{tabularx}{\linewidth}{c|ccc|ccc|ccc|ccc|ccc|ccc|ccc}
    \hline
    Blocks & \multicolumn{3}{c|}{M$_1$} & \multicolumn{3}{c|}{M$_2$} & \multicolumn{3}{c|}{M$_3$} & \multicolumn{3}{c|}{M$_4$} & \multicolumn{3}{c|}{M$_5$} & \multicolumn{3}{c|}{M$_6$} & \multicolumn{3}{c}{M$_7$} \\
    \hline
    Metric & \scriptsize{NMI} & \scriptsize{AMI} & \scriptsize{ARI} & \scriptsize{NMI} & \scriptsize{AMI} & \scriptsize{ARI} & \scriptsize{NMI} & \scriptsize{AMI} & \scriptsize{ARI} & \scriptsize{NMI} & \scriptsize{AMI} & \scriptsize{ARI} & \scriptsize{NMI} & \scriptsize{AMI} & \scriptsize{ARI} & \scriptsize{NMI} & \scriptsize{AMI} & \scriptsize{ARI} & \scriptsize{NMI} & \scriptsize{AMI} & \scriptsize{ARI} \\
    \hline
    \scriptsize{KPGNN*} & 0.39 & 0.37 & 0.07 & 0.79 & 0.78 & 0.76 & 0.76 & 0.74 & 0.58 & 0.67 & 0.64 & 0.29 & 0.73 & 0.71 & 0.47 & 0.82 & 0.79 & 0.72 & 0.55 & 0.51 & 0.12 \\ 
    \scriptsize{QSGNN*} & 0.43 & 0.41 & 0.07 & 0.81 & 0.80 & 0.77 & 0.78 & 0.76 & 0.59 & 0.71 & 0.68 & 0.29 & 0.75 & 0.73 & 0.48 & 0.83 & 0.80 & 0.73 & 0.57 & 0.54 & 0.12 \\
    \hline
    \scriptsize{SBERT*} & 0.40 & 0.38 & 0.03 & 0.86 & 0.85 & 0.73 & 0.88 & 0.87 & 0.68 & 0.82 & 0.80 & 0.36 & \underline{0.86} & \underline{0.85} & 0.61 & 0.86 & 0.83 & 0.53 & 0.64 & 0.61 & 0.09 \\
    \scriptsize{CP-Tuning*} & 0.50 & 0.48 & 0.06 & 0.87 & 0.86 & 0.73 & 0.85 & 0.84 & 0.66 & 0.79 & 0.76 & 0.27 & 0.84 & 0.83 & 0.59 & 0.87 & 0.84 & 0.58 & 0.69 & 0.67 & 0.30 \\
    \scriptsize{PromptSED*} & \underline{0.51} & \underline{0.50} & \underline{0.12} & 0.89 & 0.88 & 0.79 & 0.87 & 0.86 & 0.71 & \underline{0.83} & \underline{0.81} & 0.30 & \underline{0.86} & \underline{0.85} & 0.60 & 0.88 & 0.86 & 0.66 & \underline{0.70} & \underline{0.68} & \underline{0.31} \\
    \hline
    \scriptsize{HISEvent} & 0.38 & 0.37 & 0.09 & \underline{0.90} & \underline{0.89} & \underline{0.88} & \underline{0.90} & \underline{0.89} & \underline{0.79} & 0.77 & 0.76 & \underline{0.52} & 0.83 & 0.82 & \underline{0.63} & \underline{0.89} & \underline{0.88} & \underline{0.84} & 0.64 & 0.63 & \textbf{0.36} \\
    \hline
    \rowcolor{orange!40}
    \scriptsize{\framework{}} & \textbf{0.55} & \textbf{0.53} & \textbf{0.27} & \textbf{0.95} & \textbf{0.95} & \textbf{0.93} & \textbf{0.95} & \textbf{0.95} & \textbf{0.94} & \textbf{0.90} & \textbf{0.89} & \textbf{0.68} & \textbf{0.93} & \textbf{0.93} & \textbf{0.90} & \textbf{0.97} & \textbf{0.96} & \textbf{0.97} & \textbf{0.75} & \textbf{0.73} & 0.26 \\
     \scriptsize{Promotion} & $\uparrow$.04 & $\uparrow$.03 & $\uparrow$.15 & $\uparrow$.05 & $\uparrow$.06 & $\uparrow$.05 & $\uparrow$.05 & $\uparrow$.06 & $\uparrow$.15 & $\uparrow$.07 & $\uparrow$.08 & $\uparrow$.16 & $\uparrow$.07 & $\uparrow$.08 & $\uparrow$.27 & $\uparrow$.08 & $\uparrow$.08 & $\uparrow$.13 & $\uparrow$.05 & $\uparrow$.05 & $\downarrow$.10 \\
    \hline
    \hline
    Blocks & \multicolumn{3}{c|}{M$_8$} & \multicolumn{3}{c|}{M$_9$} & \multicolumn{3}{c|}{M$_{10}$} & \multicolumn{3}{c|}{M$_{11}$} & \multicolumn{3}{c|}{M$_{12}$} & \multicolumn{3}{c|}{M$_{13}$} & \multicolumn{3}{c}{M$_{14}$} \\
    \hline
    Metric & \scriptsize{NMI} & \scriptsize{AMI} & \scriptsize{ARI} & \scriptsize{NMI} & \scriptsize{AMI} & \scriptsize{ARI} & \scriptsize{NMI} & \scriptsize{AMI} & \scriptsize{ARI} & \scriptsize{NMI} & \scriptsize{AMI} & \scriptsize{ARI} & \scriptsize{NMI} & \scriptsize{AMI} & \scriptsize{ARI} & \scriptsize{NMI} & \scriptsize{AMI} & \scriptsize{ARI} & \scriptsize{NMI} & \scriptsize{AMI} & \scriptsize{ARI} \\
    \hline
    \scriptsize{KPGNN*} & 0.80 & 0.76 & 0.60 & 0.74 & 0.71 & 0.46 & 0.80 & 0.78 & 0.70 & 0.74 & 0.71 & 0.49 & 0.68 & 0.66 & 0.48 & 0.69 & 0.67 & 0.29 & 0.69 & 0.65 & 0.42 \\ 
    \scriptsize{QSGNN*} & 0.79 & 0.75 & 0.59 & 0.77 & 0.75 & 0.47 & 0.82 & 0.80 & 0.71 & 0.75 & 0.72 & 0.49 & 0.70 & 0.68 & 0.49 & 0.68 & 0.66 & 0.29 & 0.68 & 0.66 & 0.41 \\
    \hline
    \scriptsize{SBERT*} & \underline{0.88} & \underline{0.86} & 0.65 & 0.85 & 0.83 & 0.47 & 0.87 & 0.85 & 0.62 & 0.84 & 0.82 & 0.49 & 0.86 & 0.85 & 0.63 & 0.73 & 0.70 & 0.24 & 0.79 & 0.77 & 0.40 \\
    \scriptsize{CP-Tuning*} & 0.85 & 0.82 & 0.59 & 0.82 & 0.80 & 0.48 & 0.85 & 0.83 & 0.69 & 0.80 & 0.78 & 0.48 & 0.80 & 0.79 & 0.46 & 0.72 & 0.70 & 0.36 & 0.77 & 0.75 & 0.51 \\
    \scriptsize{PromptSED*} & 0.87 & 0.85 & 0.59 & 0.86 & 0.83 & \underline{0.56} & 0.86 & 0.85 & 0.71 & 0.84 & 0.82 & 0.51 & 0.86 & 0.85 & 0.52 & 0.74 & 0.71 & \underline{0.54} & 0.80 & 0.78 & 0.53 \\
    \hline
    \scriptsize{HISEvent} & 0.82 & 0.81 & \underline{0.68} & \underline{0.89} & \underline{0.88} & \textbf{0.65} & \underline{0.91} & \underline{0.90} & \underline{0.87} & \underline{0.85} & \underline{0.84} & \underline{0.66} & \underline{0.87} & \underline{0.87} & \underline{0.82} & \underline{0.75} & \underline{0.74} & 0.39 & \underline{0.83} & \underline{0.82} & \underline{0.71} \\
    \hline
    \rowcolor{orange!40}
    \scriptsize{\framework{}} & \textbf{0.96} & \textbf{0.95} & \textbf{0.92} & \textbf{0.91} & \textbf{0.90} & \textbf{0.65} & \textbf{0.96} & \textbf{0.96} & \textbf{0.97} & \textbf{0.95} & \textbf{0.94} & \textbf{0.96} & \textbf{0.91} & \textbf{0.91} & \textbf{0.83} & \textbf{0.88} & \textbf{0.87} & \textbf{0.77} & \textbf{0.93} & \textbf{0.92} & \textbf{0.92} \\
    \scriptsize{Promotion} & $\uparrow$.08 & $\uparrow$.09 & $\uparrow$.24 & $\uparrow$.02 & $\uparrow$.02 & - & $\uparrow$.05 & $\uparrow$.06 & $\uparrow$.10 & $\uparrow$.10 & $\uparrow$.10 & $\uparrow$.30 & $\uparrow$.04 & $\uparrow$.04 & $\uparrow$.01 & $\uparrow$.13 & $\uparrow$.13 & $\uparrow$.23 & $\uparrow$.10 & $\uparrow$.10 & $\uparrow$.21 \\
    \hline
    \hline
    Blocks & \multicolumn{3}{c|}{M$_{15}$} & \multicolumn{3}{c|}{M$_{16}$} & \multicolumn{3}{c|}{M$_{17}$} & \multicolumn{3}{c|}{M$_{18}$} & \multicolumn{3}{c|}{M$_{19}$} & \multicolumn{3}{c|}{M$_{20}$} & \multicolumn{3}{c}{M$_{21}$} \\
    \hline
    Metric & \scriptsize{NMI} & \scriptsize{AMI} & \scriptsize{ARI} & \scriptsize{NMI} & \scriptsize{AMI} & \scriptsize{ARI} & \scriptsize{NMI} & \scriptsize{AMI} & \scriptsize{ARI} & \scriptsize{NMI} & \scriptsize{AMI} & \scriptsize{ARI} & \scriptsize{NMI} & \scriptsize{AMI} & \scriptsize{ARI} & \scriptsize{NMI} & \scriptsize{AMI} & \scriptsize{ARI} & \scriptsize{NMI} & \scriptsize{AMI} & \scriptsize{ARI} \\
    \hline
    \scriptsize{KPGNN*} & 0.58 & 0.54 & 0.17 & 0.79 & 0.77 & 0.66 & 0.70 & 0.68 & 0.43 & 0.68 & 0.66 & 0.47 & 0.73 & 0.71 & 0.51 & 0.72 & 0.68 & 0.51 & 0.60 & 0.57 & 0.20 \\ 
    \scriptsize{QSGNN*} & 0.59 & 0.55 & 0.17 & 0.78 & 0.76 & 0.65 & 0.71 & 0.69 & 0.44 & 0.70 & 0.68 & 0.48 & 0.73 & 0.70 & 0.50 & 0.73 & 0.69 & 0.51 & 0.61 & 0.58 & 0.21 \\
    \hline
    \scriptsize{SBERT*} & \underline{0.70} & \underline{0.67} & 0.17 & 0.81 & 0.78 & 0.50 & 0.78 & 0.77 & 0.35 & \underline{0.82} & \underline{0.81} & 0.52 & 0.84 & 0.83 & 0.54 & 0.83 & 0.80 & 0.52 & 0.72 & 0.70 & 0.24 \\
    \scriptsize{CP-Tuning*} & 0.64 & 0.61 & 0.31 & 0.80 & 0.77 & 0.74 & 0.76 & 0.75 & 0.58 & 0.76 & 0.75 & 0.44 & 0.82 & 0.81 & 0.43 & 0.83 & 0.79 & 0.55 & 0.71 & 0.69 & 0.40 \\
    \scriptsize{PromptSED*} & 0.67 & 0.64 & \underline{0.34} & 0.82 & 0.81 & 0.75 & \underline{0.80} & \underline{0.79} & \underline{0.61} & \underline{0.82} & 0.80 & 0.49 & 0.84 & 0.83 & 0.50 & \underline{0.86} & \underline{0.83} & \underline{0.57} & \underline{0.74} & 0.72 & \underline{0.41} \\
    \hline
    \scriptsize{HISEvent} & 0.69 & \underline{0.67} & 0.27 & \underline{0.87} & \underline{0.86} & \underline{0.83} & 0.77 & 0.76 & 0.56 & 0.74 & 0.73 & \underline{0.64} & \underline{0.85} & \underline{0.84} & \underline{0.60} & 0.82 & 0.80 & \textbf{0.67} & 0.73 & \underline{0.73} & \textbf{0.46} \\
    \hline
    \rowcolor{orange!40}
    \scriptsize{\framework{}} & \textbf{0.92} & \textbf{0.91} & \textbf{0.97} & \textbf{0.92} & \textbf{0.92} & \textbf{0.86} & \textbf{0.93} & \textbf{0.93} & \textbf{0.95} & \textbf{0.87} & \textbf{0.86} & \textbf{0.69} & \textbf{0.90} & \textbf{0.89} & \textbf{0.76} & \textbf{0.87} & \textbf{0.84} & \textbf{0.67} & \textbf{0.78} & \textbf{0.76} & 0.38 \\
    \scriptsize{Promotion} & $\uparrow$.22 & $\uparrow$.24 & $\uparrow$.63 & $\uparrow$.05 & $\uparrow$.06 & $\uparrow$.03 & $\uparrow$.13 & $\uparrow$.14 & $\uparrow$.34 & $\uparrow$.05 & $\uparrow$.05 & $\uparrow$.05 & $\uparrow$.05 & $\uparrow$.05 & $\uparrow$.16 & $\uparrow$.01 & $\uparrow$.01 & - & $\uparrow$.04 & $\uparrow$.03 & $\downarrow$.08 \\
    \hline
    \end{tabularx}
    \end{threeparttable}
\end{table*}

\paragraph{Graph Construction.}
At each day $t$, given the $\mathcal{KB}^{(t)} = \{e_1,\ e_2,\ \dots \ \}$, where each event $e_i$ is represented by its keyword set $\mathrm{Keywords}_{e_i}$ and embedding $p_{e_i}$, we construct a weighted undirected graph $G_t = (V_t, E_t, W_t)$.
The node set $V_t$ consists of both the events in the current knowledge base $\mathcal{KB}^{(t)}$ and those inherited from the previous $\mathcal{KB}^{(t-1)}$ (see the inheritance mechanism below).
Formally,
\begin{equation}
\label{eq:SEE1}
V_t = 
\begin{cases}
\{e_1,e_2,\dots,e_{n_t}\}, & t=1, \\
\{e_1,e_2,\dots,e_{n_t}\} \cup \mathcal{E}_{\mathrm{inh}}^{t-1}, & t > 1,
\end{cases},
\end{equation}
where $\mathcal{E}_{\mathrm{inh}}^{t-1}$ is the set of inherited event nodes, and each node also has a keyword set and an embedding.
Edges of $G_t$ are established between event nodes sharing at least one keyword.
Formally,
\begin{equation}
\label{eq:SEE2}
E_t = \{(i,j) \ |\  i<j,\; \mathrm{Keywords}_{e_i} \cap \mathrm{Keywords}_{e_j} \neq \varnothing \},
\end{equation}
where $e_i$ and $e_j$ are events in $V_t$.
Furthermore, the weight of each edge is the cosine similarity of the embeddings of the two events it connects.
By constructing such graphs at each time $t$, we then use the structural entropy to model event evolution.

\paragraph{Structural Entropy Minimization.}
Given the constructed graph $G_t=(V_t,\ E_t,\ W_t)$, we apply structural information theory~\cite{li2016structural} to align event granularity and track event evolution.
Structural information theory encodes a graph by an encoding tree $\mathcal{T}$ that induces a partitioning $\pi_t = \{c_1,\ c_2,\ \dots\}$ of $V_t$.
The structural entropy of $G_t$
with respect to $\mathcal{T}$ is defined as
\begin{equation}
H^\mathcal{T}(G_t;\pi_t) = -\sum_{\alpha \in \mathcal{T}}\frac{g_\alpha}{vol(G_t)}\log_2\frac{vol(\alpha)}{vol(\alpha^-)},
\end{equation}
where $\alpha$ is a non-root node in $\mathcal{T}$ and $\alpha^-$ is the parent node of $\alpha$.
Structural information theory states that the $\mathcal{T}^*$
that minimizes structural entropy corresponds to the optimal partitioning.
Following~\cite{li2016structural}, we employ their structural entropy minimization algorithm to obtain the optimal partitioning of
$G_t$. 
Formally,
\begin{equation}
\pi^*_t = \{c_1^*,\ c_2^*,\ \dots\}
= \arg\min_{\pi_t \in \mathcal{P}(V_t),\ \mathcal{T}} H^{\mathcal{T}}(G_t;\pi_t),
\end{equation}
where $\mathcal{P}(V_t)$ is all partitioning of $V_t$.
Each partitioning $c_i^*$ is regarded as an aligned event $e_i^*$ at time $t$.
For each $e_i^*$, we collect the keywords of all nodes within
$c_i^*$ (excluding inherited nodes) and select the top-$15$ keywords by frequency as $\mathrm{keywords}_{e_i^*}$. 
During minimization, multiple fine-grained events of a popular event tend to be grouped into one aligned event, while sparse cold events are preserved separately.
This effect automatically aligns the event granularity with the original $\mathcal{KB}^{(t)}$.
Furthermore, aligned event in $\{e_i^*\}$ are categorized into two types: if its partitioning contains inherited node from time $t\!-\!1$, it is interpreted as the evolution of a previously existing event; otherwise, it is a newly emerging event at time $t$.

\paragraph{Inheritance Mechanism (IM)}
To continuously track event evolution over time, we design an inheritance mechanism.
Specifically, for each aligned event with its keywords
obtained at time $t\!-\!1$ through structural entropy minimization, we introduce a corresponding inherited node.
These inherited nodes participate in the subsequent graph construction and structural entropy minimization at time $t$.
Importantly, these inherited nodes are prohibited from being merged into the same partitioning during structural entropy minimization. 
This no-merge constraint ensures that each previously detected event is preserved as an independent evolving entity, even if it expands by absorbing new nodes. 
Through this inheritance mechanism, our method can effectively identify the onset of events and track the evolution of event keywords.

\paragraph{Forgetting Mechanism (FM)}
For inherited nodes participating in graph construction, we designed a forgetting mechanism to avoid outdated events from persisting as noise. 
Specifically, during the structural entropy minimization at time $t$, if an inherited
node from $t\!-\!1$ receives no support from ordinary event
nodes (i.e., no other nodes are assigned to the partitioning
of this inherited node), we regard the inherited event as outdated. 
Such events are excluded from inheritance in the next
time step $t\!+\!1$.
Through this forgetting mechanism, our method can effectively locate the end of events.

\section{Experiments}
\label{sec:experiments}
\begin{table*}[t]
  \centering
  \footnotesize
  \renewcommand{\arraystretch}{0.7}
  \setlength{\tabcolsep}{3.44pt}
  \caption{Social event detection results on Events2018. * marks results acquired with the ground truth message labels. The best results are \textbf{bolded}, the second-best results are \underline{underlined}, and the proposed method is marked in an orange background.} 
  \label{table:openset_results_event2018}
  \begin{threeparttable}
    \begin{tabularx}{\linewidth}{c|ccc|ccc|ccc|ccc|ccc|ccc|ccc|ccc}
    \hline
    Blocks & \multicolumn{3}{c|}{M$_1$} & \multicolumn{3}{c|}{M$_2$} & \multicolumn{3}{c|}{M$_3$} & \multicolumn{3}{c|}{M$_4$} & \multicolumn{3}{c|}{M$_5$} & \multicolumn{3}{c|}{M$_6$} & \multicolumn{3}{c|}{M$_7$} & \multicolumn{3}{c}{M$_8$} \\
    \hline
    Metric & \scriptsize{NMI} & \scriptsize{AMI} & \scriptsize{ARI} & \scriptsize{NMI} & \scriptsize{AMI} & \scriptsize{ARI} & \scriptsize{NMI} & \scriptsize{AMI} & \scriptsize{ARI} & \scriptsize{NMI} & \scriptsize{AMI} & \scriptsize{ARI} & \scriptsize{NMI} & \scriptsize{AMI} & \scriptsize{ARI} & \scriptsize{NMI} & \scriptsize{AMI} & \scriptsize{ARI} & \scriptsize{NMI} & \scriptsize{AMI} & \scriptsize{ARI} & \scriptsize{NMI} & \scriptsize{AMI} & \scriptsize{ARI} \\
    \hline
    \scriptsize{KPGNN*} & 0.54 & 0.54 & 0.17 & 0.56 & 0.55 & 0.18 & 0.52 & 0.55 & 0.15 & 0.55 & 0.55 & 0.17 & 0.58 & 0.57 & 0.21 & 0.59 & 0.57 & 0.21 & 0.63 & 0.61 & \underline{0.30} & 0.58 & 0.57 & 0.20 \\ 
    \scriptsize{QSGNN*} & 0.57 & 0.56 & 0.18 & 0.58 & 0.57 & 0.19 & 0.57 & 0.56 & 0.17 & 0.58 & 0.57 & 0.18 & 0.61 & 0.59 & 0.23 & 0.60 & 0.59 & 0.21 & 0.64 & 0.63 & \underline{0.30} & 0.57 & 0.55 & 0.19 \\
    \hline
    \scriptsize{SBERT*} & 0.60 & 0.60 & 0.20 & 0.62 & 0.61 & 0.29 & 0.64 & 0.63 & 0.34 & 0.61 & 0.60 & 0.23 & \underline{0.77} & \underline{0.76} & 0.47 & 0.73 & 0.73 & \underline{0.41} & 0.66 & 0.65 & 0.29 & \underline{0.76} & 0.75 & 0.50 \\
    \scriptsize{CP-Tuning*} & \underline{0.74} & \underline{0.74} & 0.36 & 0.65 & 0.64 & 0.10 & 0.57 & 0.56 & 0.31 & 0.53 & 0.52 & 0.30 & 0.62 & 0.61 & 0.35 & 0.65 & 0.64 & 0.31 & 0.50 & 0.50 & 0.11 & 0.60 & 0.59 & 0.33 \\
    \scriptsize{PromptSED*} & \textbf{0.75} & \textbf{0.75} & 0.38 & 0.66 & 0.65 & 0.28 & 0.63 & 0.62 & 0.35 & 0.58 & 0.57 & 0.34 & 0.65 & 0.63 & 0.36 & 0.68 & 0.67 & 0.38 & 0.55 & 0.54 & 0.27 & 0.69 & 0.68 & 0.39 \\
    \hline
    \scriptsize{HISEvent} & \underline{0.74} & \underline{0.74} & \textbf{0.58} & \underline{0.73} & \underline{0.73} & \underline{0.60} & \textbf{0.72} & \textbf{0.72} & \textbf{0.52} & \underline{0.67} & \underline{0.66} & \underline{0.48} & 0.74 & 0.73 & \underline{0.56} & \textbf{0.80} & \textbf{0.79} & \textbf{0.66} & \textbf{0.79} & \textbf{0.78} & \textbf{0.59} & \textbf{0.82} & \textbf{0.82} & \textbf{0.75} \\
    \hline
    \rowcolor{orange!40}
    \scriptsize{\framework{}} & 0.72 & 0.71 & \underline{0.55} & \textbf{0.77} & \textbf{0.76} & \textbf{0.68} & \underline{0.71} & \underline{0.70} & \underline{0.49} & \textbf{0.71} & \textbf{0.69} & \textbf{0.65} & \textbf{0.81} & \textbf{0.80} & \textbf{0.72} & \underline{0.78} & \underline{0.77} & \textbf{0.66} & \underline{0.72} & \underline{0.71} & \textbf{0.59} & \textbf{0.82} & \underline{0.81} & \underline{0.67} \\
    \scriptsize{Promotion} & $\downarrow$.03 & $\downarrow$.04 & $\downarrow$.03 & $\uparrow$.04 & $\uparrow$.03 & $\uparrow$.08 & $\downarrow$.01 & $\downarrow$.02 & $\downarrow$.03 & $\uparrow$.04 & $\uparrow$.03 & $\uparrow$.17 & $\uparrow$.04 & $\uparrow$.04 & $\uparrow$.16 & $\downarrow$.02 & $\downarrow$.02 & - & $\downarrow$.07 & $\downarrow$.07 & - & - & $\downarrow$.01 & $\downarrow$.08  \\
    \hline
    \hline
    Blocks & \multicolumn{3}{c|}{M$_9$} & \multicolumn{3}{c|}{M$_{10}$} & \multicolumn{3}{c|}{M$_{11}$} & \multicolumn{3}{c|}{M$_{12}$} & \multicolumn{3}{c|}{M$_{13}$} & \multicolumn{3}{c|}{M$_{14}$} & \multicolumn{3}{c|}{M$_{15}$} & \multicolumn{3}{c}{M$_{16}$} \\
    \hline
    Metric & \scriptsize{NMI} & \scriptsize{AMI} & \scriptsize{ARI} & \scriptsize{NMI} & \scriptsize{AMI} & \scriptsize{ARI} & \scriptsize{NMI} & \scriptsize{AMI} & \scriptsize{ARI} & \scriptsize{NMI} & \scriptsize{AMI} & \scriptsize{ARI} & \scriptsize{NMI} & \scriptsize{AMI} & \scriptsize{ARI} & \scriptsize{NMI} & \scriptsize{AMI} & \scriptsize{ARI} & \scriptsize{NMI} & \scriptsize{AMI} & \scriptsize{ARI} & \scriptsize{NMI} & \scriptsize{AMI} & \scriptsize{ARI} \\
    \hline
    \scriptsize{KPGNN*} & 0.48 & 0.46 & 0.10 & 0.57 & 0.56 & 0.18 & 0.54 & 0.53 & 0.16 & 0.55 & 0.56 & 0.17 & 0.60 & 0.60 & 0.28 & 0.66 & 0.65 & 0.43 & 0.60 & 0.58 & 0.25 & 0.52 & 0.50 & 0.13 \\ 
    \scriptsize{QSGNN*} & 0.52 & 0.46 & 0.13 & 0.60 & 0.58 & 0.19 & 0.60 & 0.59 & 0.20 & 0.61 & 0.59 & 0.20 & 0.59 & 0.58 & 0.27 & 0.68 & 0.67 & 0.44 & 0.63 & 0.61 & 0.27 & 0.51 & 0.50 & 0.13 \\
    \hline
    \scriptsize{SBERT*} & \underline{0.64} & \underline{0.63} & 0.23 & 0.74 & 0.72 & 0.39 & \underline{0.72} & 0.70 & 0.31 & 0.77 & 0.76 & 0.54 & 0.66 & \underline{0.65} & 0.34 & 0.69 & 0.68 & 0.43 & 0.72 & 0.71 & 0.40 & 0.66 & \underline{0.65} & 0.25 \\
    \scriptsize{CP-Tuning*} & 0.49 & 0.47 & 0.31 & 0.64 & 0.61 & 0.34 & 0.65 & 0.63 & 0.35 & 0.64 & 0.62 & 0.33 & 0.52 & 0.50 & 0.31 & 0.53 & 0.52 & 0.30 & 0.59 & 0.58 & 0.33 & 0.48 & 0.46 & 0.28 \\
    \scriptsize{PromptSED*} & 0.55 & 0.53 & 0.29 & 0.65 & 0.63 & 0.28 & 0.67 & 0.65 & 0.35 & 0.68 & 0.67 & 0.36 & 0.57 & 0.55 & 0.31 & 0.60 & 0.59 & 0.33 & 0.66 & 0.64 & 0.34 & 0.55 & 0.53 & 0.30 \\
    \hline
    \scriptsize{HISEvent} & \textbf{0.65} & \textbf{0.64} & \textbf{0.42} & \textbf{0.77} & \textbf{0.76} & \textbf{0.66} & \underline{0.72} & \underline{0.71} & \underline{0.44} & \underline{0.84} & \underline{0.83} & \underline{0.80} & \textbf{0.78} & \textbf{0.78} & \textbf{0.86} & \textbf{0.83} & \textbf{0.82} & \textbf{0.75} & \underline{0.76} & \underline{0.75} & \underline{0.61} & \textbf{0.70} & \textbf{0.69} & \underline{0.38} \\
    \hline
    \rowcolor{orange!40}
    \scriptsize{\framework{}} & \textbf{0.65} & \underline{0.63} & \underline{0.32} & \underline{0.76} & \underline{0.74} & \underline{0.57} & \textbf{0.78} & \textbf{0.76} & \textbf{0.62} & \textbf{0.87} & \textbf{0.86} & \textbf{0.83} & \underline{0.67} & \underline{0.65} & \underline{0.51} & \underline{0.73} & \underline{0.71} & \underline{0.64} & \textbf{0.79} & \textbf{0.78} & \textbf{0.70} & \underline{0.67} & 0.64 & \textbf{0.39} \\
    \scriptsize{Promotion} & - & $\downarrow$.01 & $\downarrow$.10 & $\downarrow$.01 & $\downarrow$.02 & $\downarrow$.09 & $\uparrow$.06 & $\uparrow$.05 & $\uparrow$.18 & $\uparrow$.03 & $\uparrow$.03 & $\uparrow$.03 & $\downarrow$.11 & $\downarrow$.13 & $\downarrow$.35 & $\downarrow$.10 & $\downarrow$.11 & $\downarrow$.11 & $\uparrow$.03 & $\uparrow$.03 & $\uparrow$.09 & $\downarrow$.03 & $\downarrow$.05 & $\uparrow$.01 \\
    \hline
    \end{tabularx}
    \end{threeparttable}
\end{table*}

We conduct experiments to validate the effectiveness and efficiency of the proposed~\framework{}.
In addition, we conduct ablation studies, hyperparameter studies, case studies, and visualization studies (Appendix~\ref{Appdxvi}) to demonstrate the superiority of~\framework{}.

\subsection{Experimental Setups}
\label{subsec:ES}
\subsubsection{Evaluation Metrics.}
\label{subsubsec:EM} 
To assess the effectiveness of SED, we follow previous studies~\cite{yu2025promptsed, cao2024hierarchical} and use three clustering metrics to measure the consistency between the detected event clusters and the ground truth clusters: Normalized Mutual Information (NMI)~\cite{estevez2009normalized}, Adjusted Mutual Information (AMI)~\cite{vinh2009information}, and Adjusted Rand Index (ARI)~\cite{vinh2009information}.
To assess the effectiveness of SEE, we use the popular $C_v$ metric~\cite{roder2015exploring} to evaluate the keyword coherence of evolving events and employ the Topic Diversity (TD) metric~\cite{dieng2020topic} to measure the diversity of events captured during the evolution process.
We take the average $C_v$ and TD over all time slices.
\subsubsection{Datasets.}
\label{subsubsec:D}
We conduct experiments on two public Twitter datasets: Event2012 (68,841 English messages, 503 events)~\cite{mcminn2013building} and Event2018 (64,516 French messages, 257 events)~\cite{mazoyer2020french}. Following processing in \cite{cao2024hierarchical}, the datasets are split into daily message blocks.
The detailed statistical information is presented in Appendix~\ref{AppdxDataset}.
\subsubsection{Baselines.}
\label{subsubsec:B}
To evaluate the SED performance of \framework{}, we compare it with two GNN-based methods (\textbf{KPGNN}~\cite{cao2021knowledge} and \textbf{QSGNN}~\cite{ren2022known}), three PLM-based methods (\textbf{SBERT}~\cite{reimers2019sentence}, \textbf{CP-Tuning}~\cite{xu2023making}, and \textbf{PromptSED}~\cite{yu2025promptsed}), and one structural entropy-based method (\textbf{HISEvent}~\cite{cao2024hierarchical}).
For SEE performance of \framework{}, since there are no established baselines, we compare it with topic modeling approaches~\cite{wu2024towards,wu2024survey} that can also extract event keywords. These include three non-dynamic models (\textbf{ProdLDA}~\cite{srivastava2017autoencoding}, \textbf{DecTM}~\cite{wu2021discovering}, and \textbf{TSCTM}~\cite{wu2022mitigating}) and two dynamic models (\textbf{CFDTM}~\cite{wu2024modeling} and \textbf{BERTopic}~\cite{grootendorst2022bertopic}).
Appendix~\ref{AppdxID} and~\ref{AppdxBaselines} show additional descriptions.

\subsection{Main Results}
\label{subsec:MR}
\subsubsection{Effectiveness Analysis of Unsupervised SED}
\label{subsub:EAUSED}
The SED evaluation results of \framework{} are reported in Tables~\ref{table:openset_results_event2012} and~\ref{table:openset_results_event2018}, where both Detection-LLM and Evaluation-LLM are instantiated with deepseek-r1:32b. On the English dataset, \framework{} consistently outperforms all baselines across nearly all message blocks, achieving maximum gains of 0.22 in NMI, 0.24 in AMI, and 0.63 in ARI (e.g., block M$_{15}$). These improvements highlight the effectiveness of global event guidance in forming more accurate event clusters. Importantly, \framework{} maintains stable promotion in both dense message blocks (e.g., M$_1$, M$_7$, M$_{12}$) and sparse message blocks (e.g., M$_{16}$, M$_{20}$), demonstrating robustness to varying data distributions.
On the French dataset, \framework{} achieves the best or second-best performance in almost all blocks. The observed marginal decrease is due to the use of a relatively small 32B LLM, which has limited capacity for French comprehension. 
In practical deployment, selecting language-specific LLMs can further enhance performance. Furthermore, beyond strong SED performance, \framework{} uniquely supports SEE, which none of the baseline methods provide.
\subsubsection{Effectiveness Analysis of Unsupervised SEE}
\label{subsub:EAUSEE}
\begin{table}[t]
    \setlength{\tabcolsep}{3.48pt}
    \centering
    \footnotesize
    \renewcommand{\arraystretch}{1}
    \caption{Social event Evolution results. In dynamic methods, the best results are bolded, the second-best are \underline{underlined}.\vspace{-2mm}}
    \label{table:SEE result}        
    \begin{tabularx}{\linewidth}{c|cccc|cccc}
        \hline
         & \multicolumn{4}{c|}{Events2012} & \multicolumn{4}{c}{Events2018} \\
        \hline
        Method & $C_v$ & TD & Avg & Dynamic & $C_v$ & TD & Avg & Dynamic \\
        \hline
        ProdLDA & 0.46 & 0.48 & 0.47 & $\times$ & 0.40 & 0.74 & 0.57 & $\times$ \\
        DecTM & 0.48 & 0.63 & 0.56 & $\times$ & 0.39 & 0.85 & 0.62 & $\times$ \\
        TSCTM & 0.44 & 0.93 & 0.69 & $\times$ & 0.36 & 0.96 & 0.66 & $\times$ \\
        \hline
        CFDTM & \underline{0.65} & 0.14 & 0.40 & $\checkmark$ & \underline{0.66} & 0.34 & 0.50 & $\checkmark$ \\
        Bertopic & 0.61 & \underline{0.42} & \underline{0.52} & $\checkmark$ & 0.59 & 0.39 & 0.49 & $\checkmark$ \\
        \hline
        \framework{} w/o IM & 0.53 & 0.90 & 0.72 & $\times$ & 0.53 & 0.87 & 0.70 & $\times$ \\
        \framework{} w/o FM & \textbf{0.76} & 0.28 & \underline{0.52} & $\checkmark$ & \textbf{0.69} & \underline{0.42} & \underline{0.56} & $\checkmark$ \\
        \rowcolor{orange!40}
        \framework{} & 0.49 & \textbf{0.88} & \textbf{0.69} & $\checkmark$ & 0.52 & \textbf{0.87} & \textbf{0.70} & $\checkmark$ \\
        \hline
    \end{tabularx}
    \vspace{-2mm}
\end{table}

\begin{table}[t]
    \setlength{\tabcolsep}{1.73pt}
    \centering
    \footnotesize
    \renewcommand{\arraystretch}{1}
    \caption{Ablation results for SED on two datasets. The best results are bolded, the second-best are \underline{underlined}.\vspace{-2mm}}
    \label{table:Ab result}        
    \begin{tabularx}{\linewidth}{c|ccc|ccc|ccc|ccc}
    \hline
    Blocks & \multicolumn{3}{c|}{\tiny{M$_3$(Event2012)}} & \multicolumn{3}{c|}{\tiny{M$_{10}$(Event2012)}} & \multicolumn{3}{c|}{\tiny{M$_8$(Event2018)}} & \multicolumn{3}{c}{\tiny{M$_{15}$(Event2018)}} \\
    \hline
    Metric & \scriptsize{NMI} & \scriptsize{AMI} & \scriptsize{ARI} & \scriptsize{NMI} & \scriptsize{AMI} & \scriptsize{ARI} & \scriptsize{NMI} & \scriptsize{AMI} & \scriptsize{ARI} & \scriptsize{NMI} & \scriptsize{AMI} & \scriptsize{ARI} \\
    \hline
    \scriptsize{\framework{} w/o KMS} & .84 & .81 & .61 & .93 & .92 & \underline{.88} & .72 & .67 & .41 & .73 & .69 & .55 \\
    \scriptsize{\framework{} w/o $\mathcal{KB}$} & .86 & \underline{.85} & .57 & .91 & .89 & .72 & .73 & .68 & .34 & .68 & .64 & .28 \\
    \scriptsize{\framework{} w/o KBM} & \underline{.95} & \textbf{.95} & \underline{.94} & \underline{.96} & \underline{.96} & \textbf{.97} & \underline{.81} & .79 & \underline{.67} & .77 & .76 & .68 \\
    \hline
    \rowcolor{orange!40}
    \scriptsize{\framework{} (deepseek-r1:32b)} & \underline{.95} & \textbf{.95} & \underline{.94} & \underline{.96} & \underline{.96} & \textbf{.97} & \textbf{.82} & \textbf{.81} & \underline{.67} & \underline{.79} & \underline{.78} & .70 \\
    \scriptsize{\framework{} + deepseek-r1:70b} & \textbf{.96} & \textbf{.95} & \textbf{.95} & \underline{.96} & \underline{.96} & \textbf{.97} & \underline{.81} & .79 & .64 & \underline{.79} & \underline{.78} & \underline{.72} \\
    \scriptsize{\framework{} + GPT-4o-mini} & \textbf{.96} & \textbf{.95} & \textbf{.95} & \textbf{.97} & \textbf{.97} & \textbf{.97} & \textbf{.82} & \textbf{.81} & \textbf{.68} & \textbf{.81} & \textbf{.80} & \textbf{.77} \\
    \scriptsize{\framework{} + GPT-4o} & \textbf{.96} & \textbf{.95} & \textbf{.95} & \textbf{.97} & \underline{.96} & \textbf{.97} & \textbf{.82} & \underline{.80} & \textbf{.68} & \underline{.79} & \underline{.78} & .68 \\
    \hline
    \end{tabularx}
    \vspace{-5mm}
\end{table}

The SEE evaluation results of \framework{} are reported in Table~\ref{table:SEE result}, and all values are with two decimal precision. 
For non-dynamic methods, we execute the algorithms independently on each daily block.
When a method achieves a high $C_v$ but a very low TD value, the keywords for different events exhibit significant overlap, meaning these events are actually the same real-world event (e.g., CFDTM, Bertopic).
When a method achieves a very high TD but a low $C_v$, it excessively punishes keyword sharing across events, resulting in incoherent keywords that fail to adequately describe the events(e.g., TSCTM).
The results reveal that \framework{} achieves a better balance between $C_v$ and TD, with its average score consistently surpassing all dynamic baselines. 
This advantage stems from maintaining a high-quality knowledge base during the SED process and from using structural entropy minimization to align event granularity. 
Moreover, unlike non-dynamic approaches that fail in real-world applications, \framework{} can effectively track event evolution over time.

\subsection{Ablation Study}
\label{subsec:AS}
We conduct ablation studies on both datasets to further demonstrate the effectiveness of each module in \framework{}. 
For SED, as shown in Table~\ref{table:Ab result}, removing the KMS, $\mathcal{KB}$, or KBM modules leads to varying degrees of performance degradation. 
This is because the KMS module samples key messages instead of all messages, thereby improving the quality of LLM queries; the KB module provides event information as global guidance, enhancing the LLM’s capability; and the KBM module maintains the freshness of the knowledge base, which becomes particularly important for larger message blocks (e.g., M$_8$ and M$_{15}$ in Events2018).
For SEE, the results in Table~\ref{table:SEE result} demonstrate that removing the IM module prevents \framework{} from capturing the dynamic evolution of events, whereas removing the FM module results in insufficient event diversity, confirming the necessity of both modules.
Finally, we ablate \framework{} with different base LLMs. 
While larger LLMs naturally achieve higher accuracy, \framework{} with a 32B LLM already delivers competitive performance while significantly reducing computational resources.

\subsection{Hyperparameter Study}
\label{subsec:HS}
\begin{figure}[t]
    \centering
    \subfigtopskip=2pt 
    \subfigbottomskip=2pt 
    \subfigcapskip=-4pt 
    \subfigure[M3 (Event2012)]
    {
        \label{level.sub.1}
        \includegraphics[width=0.4\linewidth,trim={0cm 23 0cm 48}, clip]{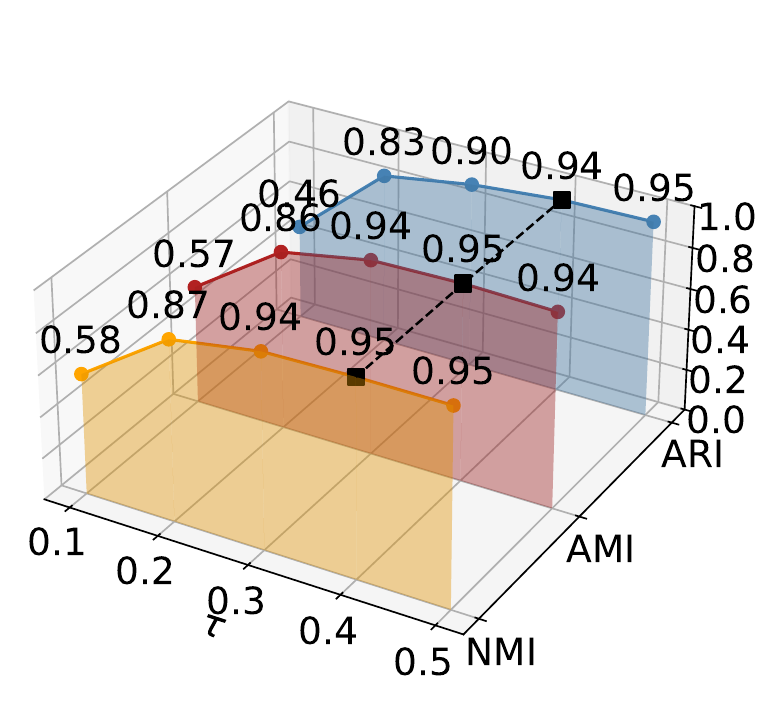}
    }
    \quad 
    \subfigure[M10 (Event2012)]
    {
        \label{level.sub.2}
        \includegraphics[width=0.4\linewidth,trim={0cm 23 0cm 48}, clip]{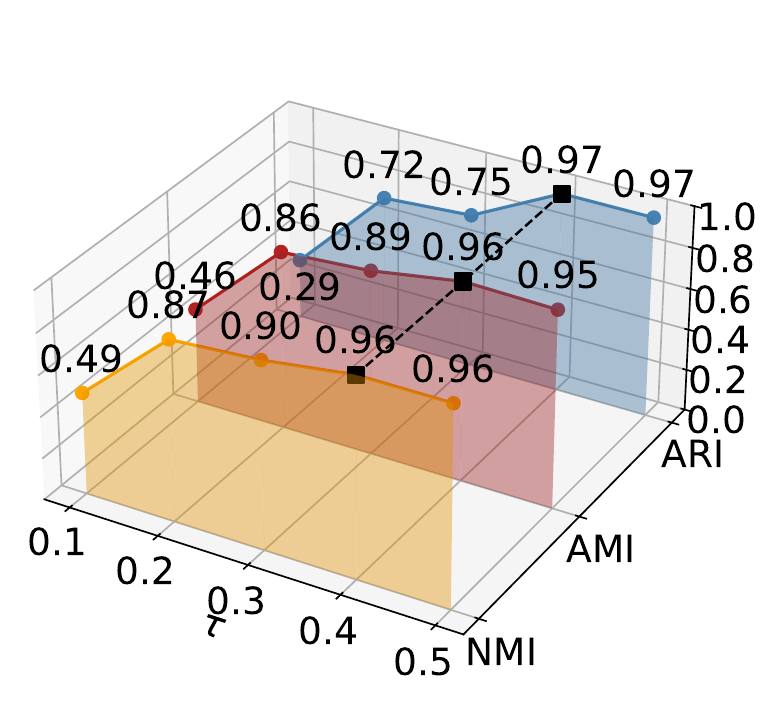}
    }
    \\
    \subfigure[M8 (Event2018)]
    {
        \label{level.sub.3}
        \includegraphics[width=0.4\linewidth,trim={0cm 23 0cm 48}, clip]{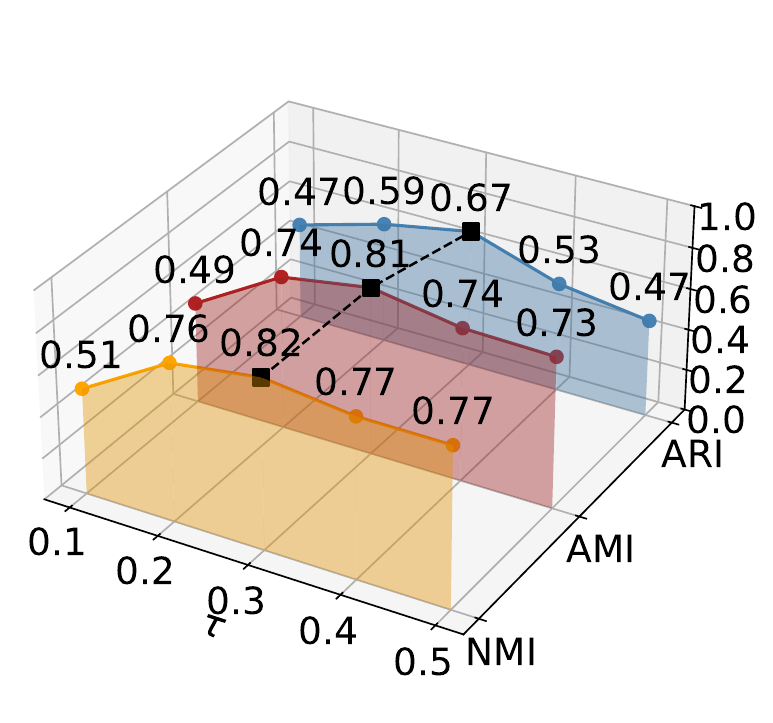}
    }
    \quad
    \subfigure[M15 (Event2018)]
    {
        \label{level.sub.4}
        \includegraphics[width=0.4\linewidth,trim={0cm 23 0cm 48}, clip]{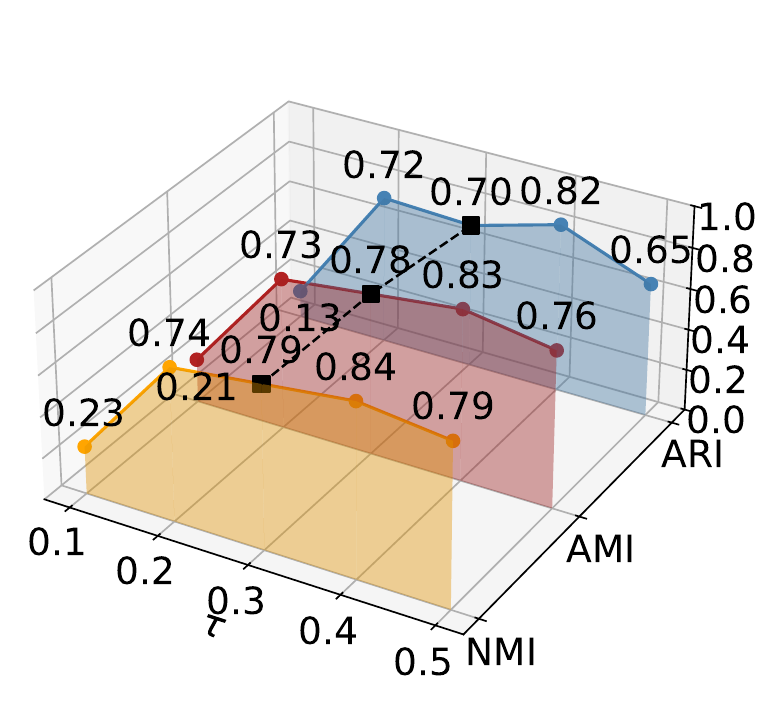}
    }
    \vspace{-0.35cm}
    \caption{Sensitivity of hyperparameter $\tau$ on four blocks.}
    \label{fic: hyperparameter}
\end{figure}

\begin{figure}[h]
    \centering
    \includegraphics[width=1\linewidth]{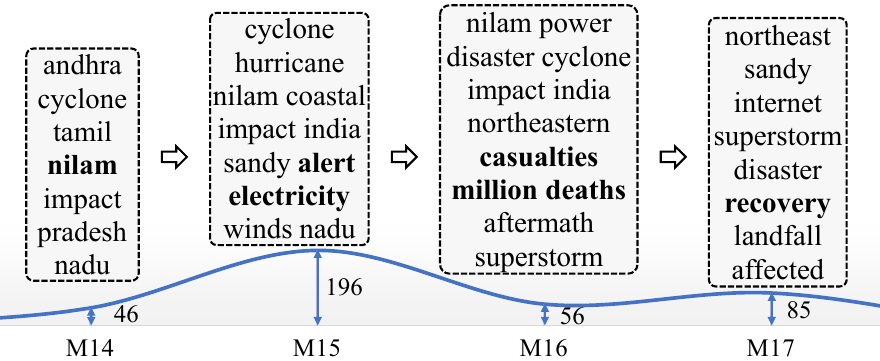}
    \vspace{-4mm}
    \caption{Case study for SEE of \framework{}. Evolution keywords of an Event about Cyclone Nilam from the Event2012 dataset.\vspace{-2mm}}
    \label{fic: case}
\end{figure}

We conduct a hyperparameter study of the similarity threshold $\tau$ in the key message sampling module (Section~\ref{subsubsec:AC}), with the results shown in Figure~\ref{fic: hyperparameter}. 
Figure~\ref{fic: hyperparameter} indicates that setting $\tau$ too low (e.g., $\tau=0.1$) significantly decreases detection performance across all blocks, as messages from different events are aggregated into the same anchor and cannot be further distinguished during detection. 
As $\tau$ increases, performance improves, with the best results observed around $\tau=0.4$ on the Event2012 and $\tau=0.3$ on the Event2018.
\framework{} remains relatively robust to $\tau$ within this reasonable range. 
However, when $\tau$ is set too high, detection efficiency decreases. 
Therefore, selecting an appropriate $\tau$ enables a favorable balance between performance and efficiency.

\vspace{-1mm}
\subsection{Case Study}
\label{subsec:CS}
We conduct a case study to demonstrate \framework{}’s capability in SEE. 
Figure~\ref{fic: case} presents the evolution keywords obtained by \framework{} for the detected event Cyclone Nilam\footnote{https://en.wikipedia.org/wiki/Cyclone\_Nilam} over a four-day period.
The curve in the figure shows changes in the event's discussion intensity.
The bold keywords in the figure~\ref{fic: case} indicate that \framework{} not only correctly localizes the Nilam region on the first day, but also continuously tracks the impact and damages of the disaster in the following days (e.g., electrical issue in M$_{15}$ and casualties in M$_{16}$). 
Finally, \framework{} captures the post-disaster recovery process. 
This ability to consistently track events across time periods demonstrates the effectiveness of the inheritance and forgetting mechanisms in the evolution framework \framework{}.

\vspace{-1mm}
\subsection{Efficiency Analysis}
\label{subsec:EA}
\begin{figure}[t]
    \centering
    \includegraphics[width=1\linewidth]{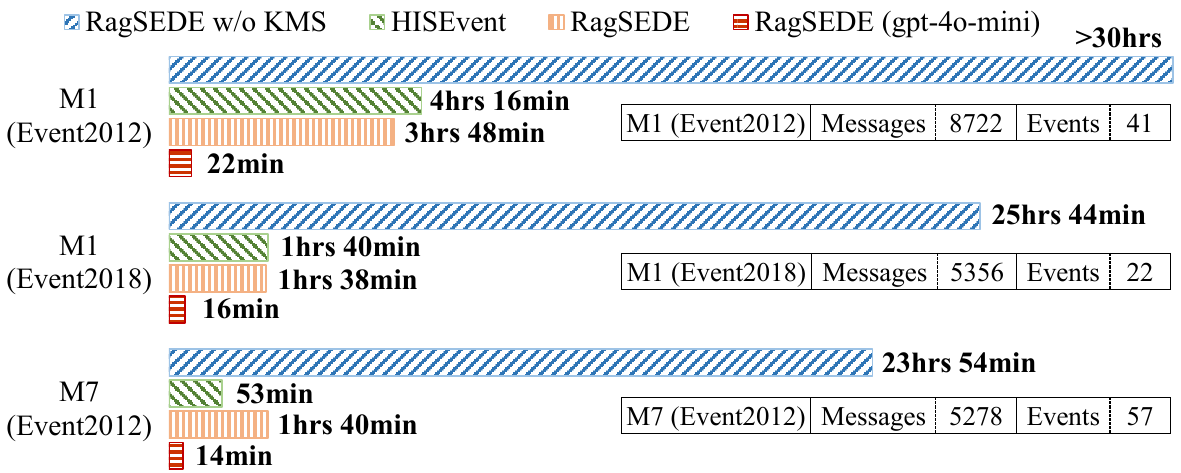}
    \caption{Running time comparison on the largest blocks.\vspace{-4mm}}
    \label{fic: efficiency}
\end{figure}

We report the running time of the proposed \framework{}, \framework{} without the key message sampling module, and the strongest baseline HISEvent on the three largest message blocks of datasets, as shown in Figure~\ref{fic: efficiency}. 
The results demonstrate that incorporating the KMS module improves the efficiency of \framework{} by up to 15 times, as it effectively avoids redundant LLM queries on semantically similar messages. 
Furthermore, compared to HISEvent, which incurs higher computational complexity, \framework{} achieves superior efficiency on large-scale message blocks that more closely reflect real-world scenarios.
Importantly, \framework{} employs the locally deployed deepseek-r1:32b, constrained by our local computational resources. 
When the backbone LLM is replaced with gpt-4o-mini, we observe a further improvement in efficiency.

\section{Related Work}
\label{sec:related_work}
In the past decade, SED~\cite{atefeh2015survey} has been extensively studied. 
Early approaches relied on incremental clustering~\cite{aggarwal2012event, hu2017adaptive}, topic modeling~\cite{zhou2014event, xing2016hashtag, wang2016using}, and community detection~\cite{fedoryszak2019real, liu2020story, liu2018event} to identify events from social streams. 
While effective to some extent, these methods often suffer from limitations in scalability, adaptability, and semantic expressiveness.
With the emergence of GNNs, which are capable of modeling complex structural dependencies among messages, GNN-based methods have rapidly become the dominant approach for SED.
These methods span a wide spectrum, including supervised approaches based on evidential learning and contrastive learning~\cite{ren2022evidential, ren2023uncertainty, ren2022known, cao2021knowledge}, as well as unsupervised~\cite{guo2024unsupervised}, cross-lingual~\cite{ren2024toward}, and multimodal~\cite{zhao2017real} approaches. 
More recently, the rapid advancement of PLMs has enabled the use of their strong contextual reasoning capabilities in SED tasks.
\cite{li2024relational, yu2025promptsed} utilizes PLMs and appropriate prompts to obtain message embeddings, followed by clustering to obtain events.
In addition, structural entropy~\cite{li2016structural, peng2025adaptive}, as an unsupervised clustering method, has shown remarkable potential for social event analysis~\cite{cao2024hierarchical}.
However, none of the above methods can track the evolution of detected events, and our method achieves this, which is our advantage.

\section{Conclusion}
\label{sec:conclusion}
In this paper, we propose \framework{}, a novel unsupervised model for SED and SEE. 
First, \framework{} introduces a key message sampling module that improves the quality of LLM queries while substantially reducing computational overhead. 
Second, \framework{} proposes a new RAG-based event detection paradigm that leverages a knowledge base for global semantic guidance. 
Finally, \framework{} leverages structural entropy to track the evolution of events over time. 
Extensive experiments on two benchmark datasets demonstrate that \framework{} outperforms all baselines. 
In future work, we plan to extend this framework to multimodal scenes.


\begin{acks}
This work is supported by the NSFC through grants U25B2029, 62322202, 62441612, and 62432006, Beijing Natural Science Foundation through grant L253021, the Pioneer and Leading Goose R\&D Program of Zhejiang through grant 2025C02044, Local Science and Technology Development Fund of Hebei Province Guided by the Central Government of China through grant 254Z9902G, Hebei Natural Science Foundation through grant F2024210008, Major Science Technology Special Projects of Yunnan Province through grants 202502AD080012 and 202502AD080006, and the Fundamental Research Funds for the Central Universities.
\end{acks}

\bibliographystyle{ACM-Reference-Format}
\bibliography{7_references}

@String{Computer = "{IEEE} Computer" }

@String{Springer = "Springer-Verlag" }

@inproceedings{qian2023open,
  title={Open-world social event classification},
  author={Qian, Shengsheng and Chen, Hong and Xue, Dizhan and Fang, Quan and Xu, Changsheng},
  booktitle={Proceedings of the ACM web conference 2023},
  pages={1562--1571},
  year={2023}
}

@inproceedings{xian2025community,
  title={Community detection in large-scale complex networks via structural entropy game},
  author={Xian, Yantuan and Li, Pu and Peng, Hao and Yu, Zhengtao and Xiang, Yan and Yu, Philip S},
  booktitle={Proceedings of the ACM on Web Conference 2025},
  pages={3930--3941},
  year={2025}
}

@inproceedings{cao2021knowledge,
  title={Knowledge-preserving incremental social event detection via heterogeneous gnns},
  author={Cao, Yuwei and Peng, Hao and Wu, Jia and Dou, Yingtong and Li, Jianxin and Yu, Philip S},
  booktitle={Proceedings of the web conference 2021},
  pages={3383--3395},
  year={2021}
}

@article{peng2021streaming,
  title={Streaming social event detection and evolution discovery in heterogeneous information networks},
  author={Peng, Hao and Li, Jianxin and Song, Yangqiu and Yang, Renyu and Ranjan, Rajiv and Yu, Philip S and He, Lifang},
  journal={ACM Transactions on Knowledge Discovery from Data (TKDD)},
  volume={15},
  number={5},
  pages={1--33},
  year={2021},
  publisher={ACM New York, NY, USA}
}

@article{fortunato202220,
  title={20 years of network community detection},
  author={Fortunato, Santo and Newman, Mark EJ},
  journal={Nature Physics},
  volume={18},
  number={8},
  pages={848--850},
  year={2022},
  publisher={Nature Publishing Group UK London}
}

@article{zhao2024recommender,
  title={Recommender systems in the era of large language models (llms)},
  author={Zhao, Zihuai and Fan, Wenqi and Li, Jiatong and Liu, Yunqing and Mei, Xiaowei and Wang, Yiqi and Wen, Zhen and Wang, Fei and Zhao, Xiangyu and Tang, Jiliang and others},
  journal={IEEE Transactions on Knowledge and Data Engineering},
  volume={36},
  number={11},
  pages={6889--6907},
  year={2024},
  publisher={IEEE}
}

@article{qian2022integrating,
  title={Integrating multi-label contrastive learning with dual adversarial graph neural networks for cross-modal retrieval},
  author={Qian, Shengsheng and Xue, Dizhan and Fang, Quan and Xu, Changsheng},
  journal={IEEE Transactions on Pattern Analysis and Machine Intelligence},
  volume={45},
  number={4},
  pages={4794--4811},
  year={2022},
  publisher={IEEE}
}

@article{rajput2023recommender,
  title={Recommender systems with generative retrieval},
  author={Rajput, Shashank and Mehta, Nikhil and Singh, Anima and Hulikal Keshavan, Raghunandan and Vu, Trung and Heldt, Lukasz and Hong, Lichan and Tay, Yi and Tran, Vinh and Samost, Jonah and others},
  journal={Advances in Neural Information Processing Systems},
  volume={36},
  pages={10299--10315},
  year={2023}
}

@inproceedings{dai2024bias,
  title={Bias and unfairness in information retrieval systems: New challenges in the llm era},
  author={Dai, Sunhao and Xu, Chen and Xu, Shicheng and Pang, Liang and Dong, Zhenhua and Xu, Jun},
  booktitle={Proceedings of the 30th ACM SIGKDD Conference on Knowledge Discovery and Data Mining},
  pages={6437--6447},
  year={2024}
}

@article{yu2025promptsed,
  title={PromptSED: An evolving topic-enhanced prompting framework for incremental social event detection},
  author={Yu, Xiaoyan and Ren, Jiaqian and Jiang, Lei and Peng, Hao and Hao, Zhifeng and Sun, Li and Peng, Kun and Zhu, Liehuang and Yu, Philip S},
  journal={Neural Networks},
  pages={107772},
  year={2025},
  publisher={Elsevier}
}

@article{cui2021mvgan,
  title={MVGAN: Multi-view graph attention network for social event detection},
  author={Cui, Wanqiu and Du, Junping and Wang, Dawei and Kou, Feifei and Xue, Zhe},
  journal={ACM Transactions on Intelligent Systems and Technology (TIST)},
  volume={12},
  number={3},
  pages={1--24},
  year={2021},
  publisher={ACM New York, NY}
}

@inproceedings{cao2024hierarchical,
  title={Hierarchical and incremental structural entropy minimization for unsupervised social event detection},
  author={Cao, Yuwei and Peng, Hao and Yu, Zhengtao and Yu, Philip S},
  booktitle={Proceedings of the AAAI conference on artificial intelligence},
  volume={38},
  number={8},
  pages={8255--8264},
  year={2024}
}

@article{li2016structural,
  title={Structural information and dynamical complexity of networks},
  author={Li, Angsheng and Pan, Yicheng},
  journal={IEEE Transactions on Information Theory},
  volume={62},
  number={6},
  pages={3290--3339},
  year={2016},
  publisher={IEEE}
}

@article{li2024relational,
  title={Relational prompt-based pre-trained language models for social event detection},
  author={Li, Pu and Yu, Xiaoyan and Peng, Hao and Xian, Yantuan and Wang, Linqin and Sun, Li and Zhang, Jingyun and Yu, Philip S},
  journal={ACM Transactions on Information Systems},
  volume={43},
  number={1},
  pages={1--43},
  year={2024},
  publisher={ACM New York, NY}
}

@inproceedings{ren2022known,
  title={From known to unknown: Quality-aware self-improving graph neural network for open set social event detection},
  author={Ren, Jiaqian and Jiang, Lei and Peng, Hao and Cao, Yuwei and Wu, Jia and Yu, Philip S and He, Lifang},
  booktitle={Proceedings of the 31st ACM International Conference on Information \& Knowledge Management},
  pages={1696--1705},
  year={2022}
}

@article{peng2022reinforced,
  title={Reinforced, incremental and cross-lingual event detection from social messages},
  author={Peng, Hao and Zhang, Ruitong and Li, Shaoning and Cao, Yuwei and Pan, Shirui and Yu, Philip S},
  journal={IEEE Transactions on Pattern Analysis and Machine Intelligence},
  volume={45},
  number={1},
  pages={980--998},
  year={2022},
  publisher={IEEE}
}

@inproceedings{peng2019fine,
  title={Fine-grained event categorization with heterogeneous graph convolutional networks},
  author={Peng, Hao and Li, Jianxin and Gong, Qiran and Song, Yangqiu and Ning, Yuanxing and Lai, Kunfeng and Yu, Philip S},
  booktitle={Proceedings of the 28th International Joint Conference on Artificial Intelligence},
  pages={3238--3245},
  year={2019}
}

@inproceedings{devlin2019bert,
  title={Bert: Pre-training of deep bidirectional transformers for language understanding},
  author={Devlin, Jacob and Chang, Ming-Wei and Lee, Kenton and Toutanova, Kristina},
  booktitle={Proceedings of the 2019 conference of the North American chapter of the association for computational linguistics: human language technologies, volume 1 (long and short papers)},
  pages={4171--4186},
  year={2019}
}

@inproceedings{reimers2019sentence,
  title={Sentence-BERT: Sentence Embeddings using Siamese BERT-Networks},
  author={Reimers, Nils and Gurevych, Iryna},
  booktitle={Proceedings of the 2019 Conference on Empirical Methods in Natural Language Processing and the 9th International Joint Conference on Natural Language Processing (EMNLP-IJCNLP)},
  pages={3982--3992},
  year={2019}
}

@inproceedings{ester1996density,
  title={A density-based algorithm for discovering clusters in large spatial databases with noise},
  author={Ester, Martin and Kriegel, Hans-Peter and Sander, J{\"o}rg and Xu, Xiaowei and others},
  booktitle={kdd},
  volume={96},
  number={34},
  pages={226--231},
  year={1996}
}

@inproceedings{yu2025towards,
  title={Towards effective, efficient and unsupervised social event detection in the hyperbolic space},
  author={Yu, Xiaoyan and Wei, Yifan and Zhou, Shuaishuai and Yang, Zhiwei and Sun, Li and Peng, Hao and Zhu, Liehuang and Yu, Philip S},
  booktitle={Proceedings of the AAAI Conference on Artificial Intelligence},
  volume={39},
  number={12},
  pages={13106--13114},
  year={2025}
}

@inproceedings{knights2009detecting,
  title={Detecting topic drift with compound topic models},
  author={Knights, Dan and Mozer, Michael and Nicolov, Nicolas},
  booktitle={Proceedings of the International AAAI Conference on Web and Social Media},
  volume={3},
  number={1},
  pages={242--245},
  year={2009}
}

@article{liu2020event,
  title={Event detection and evolution in multi-lingual social streams},
  author={Liu, Yaopeng and Peng, Hao and Li, Jianxin and Song, Yangqiu and Li, Xiong},
  journal={Frontiers of Computer Science},
  volume={14},
  number={5},
  pages={145612},
  year={2020},
  publisher={Springer}
}

@article{lewis2020retrieval,
  title={Retrieval-augmented generation for knowledge-intensive nlp tasks},
  author={Lewis, Patrick and Perez, Ethan and Piktus, Aleksandra and Petroni, Fabio and Karpukhin, Vladimir and Goyal, Naman and K{\"u}ttler, Heinrich and Lewis, Mike and Yih, Wen-tau and Rockt{\"a}schel, Tim and others},
  journal={Advances in neural information processing systems},
  volume={33},
  pages={9459--9474},
  year={2020}
}

@inproceedings{ma2025enhanced,
  title={Enhanced Social Event Detection through Dynamically Weighted Meta-Paths Modeling},
  author={Ma, Congbo and Qiu, Zitai and Wang, Hu and Du, Jing and Xue, Shan and Wu, Jia and Yang, Jian},
  booktitle={Companion Proceedings of the ACM on Web Conference 2025},
  pages={1184--1188},
  year={2025}
}

@article{estevez2009normalized,
  title={Normalized mutual information feature selection},
  author={Est{\'e}vez, Pablo A and Tesmer, Michel and Perez, Claudio A and Zurada, Jacek M},
  journal={IEEE Transactions on neural networks},
  volume={20},
  number={2},
  pages={189--201},
  year={2009},
  publisher={IEEE}
}

@inproceedings{vinh2009information,
  title={Information theoretic measures for clusterings comparison: is a correction for chance necessary?},
  author={Vinh, Nguyen Xuan and Epps, Julien and Bailey, James},
  booktitle={Proceedings of the 26th annual international conference on machine learning},
  pages={1073--1080},
  year={2009}
}

@inproceedings{roder2015exploring,
  title={Exploring the space of topic coherence measures},
  author={R{\"o}der, Michael and Both, Andreas and Hinneburg, Alexander},
  booktitle={Proceedings of the eighth ACM international conference on Web search and data mining},
  pages={399--408},
  year={2015}
}

@article{dieng2020topic,
  title={Topic modeling in embedding spaces},
  author={Dieng, Adji B and Ruiz, Francisco JR and Blei, David M},
  journal={Transactions of the Association for Computational Linguistics},
  volume={8},
  pages={439--453},
  year={2020},
  publisher={MIT Press One Rogers Street, Cambridge, MA 02142-1209, USA journals-info~…}
}

@inproceedings{mcminn2013building,
  title={Building a large-scale corpus for evaluating event detection on twitter},
  author={McMinn, Andrew J and Moshfeghi, Yashar and Jose, Joemon M},
  booktitle={Proceedings of the 22nd ACM international conference on Information \& Knowledge Management},
  pages={409--418},
  year={2013}
}

@inproceedings{mazoyer2020french,
  title={A french corpus for event detection on twitter},
  author={Mazoyer, B{\'e}atrice and Cag{\'e}, Julia and Herv{\'e}, Nicolas and Hudelot, C{\'e}line},
  booktitle={Twelfth Language Resources and Evaluation Conference},
  pages={6220--6227},
  year={2020},
  organization={European Language Resources Association (ELRA)}
}

@inproceedings{xu2023making,
  title={Making pre-trained language models end-to-end few-shot learners with contrastive prompt tuning},
  author={Xu, Ziyun and Wang, Chengyu and Qiu, Minghui and Luo, Fuli and Xu, Runxin and Huang, Songfang and Huang, Jun},
  booktitle={Proceedings of the sixteenth ACM international conference on web search and data mining},
  pages={438--446},
  year={2023}
}

@inproceedings{srivastava2017autoencoding,
  title={Autoencoding Variational Inference for Topic Models},
  author={Srivastava, Akash and Sutton, Charles},
  booktitle={5th International Conference on Learning Representations},
  year={2017}
}

@inproceedings{wu2021discovering,
  title={Discovering topics in long-tailed corpora with causal intervention},
  author={Wu, Xiaobao and Li, Chunping and Miao, Yishu},
  booktitle={Findings of the Association for Computational Linguistics: ACL-IJCNLP 2021},
  pages={175--185},
  year={2021}
}

@inproceedings{wu2022mitigating,
  title={Mitigating Data Sparsity for Short Text Topic Modeling by Topic-Semantic Contrastive Learning},
  author={Wu, Xiaobao and Tuan, Luu Anh and Dong, Xinshuai},
  booktitle={Proceedings of the 2022 Conference on Empirical Methods in Natural Language Processing},
  pages={2748--2760},
  year={2022}
}

@inproceedings{wu2024modeling,
  title={Modeling Dynamic Topics in Chain-Free Fashion by Evolution-Tracking Contrastive Learning and Unassociated Word Exclusion},
  author={Wu, Xiaobao and Dong, Xinshuai and Pan, Liangming and Nguyen, Thong and Tuan, Luu Anh},
  booktitle={Findings of the Association for Computational Linguistics ACL 2024},
  pages={3088--3105},
  year={2024}
}

@article{grootendorst2022bertopic,
  title={BERTopic: Neural topic modeling with a class-based TF-IDF procedure},
  author={Grootendorst, Maarten},
  journal={arXiv preprint arXiv:2203.05794},
  year={2022}
}

@inproceedings{wu2024towards,
  title={Towards the TopMost: A Topic Modeling System Toolkit},
  author={Wu, Xiaobao and Pan, Fengjun and Tuan, Luu Anh},
  booktitle={Proceedings of the 62nd Annual Meeting of the Association for Computational Linguistics (Volume 3: System Demonstrations)},
  pages={31--41},
  year={2024}
}

@article{wu2024survey,
  title={A survey on neural topic models: methods, applications, and challenges},
  author={Wu, Xiaobao and Nguyen, Thong and Luu, Anh Tuan},
  journal={Artificial Intelligence Review},
  volume={57},
  number={2},
  pages={18},
  year={2024},
  publisher={Springer}
}

@inproceedings{shi2024generate,
  title={Generate-then-Ground in Retrieval-Augmented Generation for Multi-hop Question Answering},
  author={Shi, Zhengliang and Zhang, Shuo and Sun, Weiwei and Gao, Shen and Ren, Pengjie and Chen, Zhumin and Ren, Zhaochun},
  booktitle={Proceedings of the 62nd Annual Meeting of the Association for Computational Linguistics (Volume 1: Long Papers)},
  pages={7339--7353},
  year={2024}
}

@inproceedings{fan2024survey,
  title={A survey on rag meeting llms: Towards retrieval-augmented large language models},
  author={Fan, Wenqi and Ding, Yujuan and Ning, Liangbo and Wang, Shijie and Li, Hengyun and Yin, Dawei and Chua, Tat-Seng and Li, Qing},
  booktitle={Proceedings of the 30th ACM SIGKDD conference on knowledge discovery and data mining},
  pages={6491--6501},
  year={2024}
}

@article{zhao2024retrieval,
  title={Retrieval-augmented generation for ai-generated content: A survey},
  author={Zhao, Penghao and Zhang, Hailin and Yu, Qinhan and Wang, Zhengren and Geng, Yunteng and Fu, Fangcheng and Yang, Ling and Zhang, Wentao and Jiang, Jie and Cui, Bin},
  journal={arXiv preprint arXiv:2402.19473},
  year={2024}
}

@article{atefeh2015survey,
  title={A survey of techniques for event detection in twitter},
  author={Atefeh, Farzindar and Khreich, Wael},
  journal={Computational Intelligence},
  volume={31},
  number={1},
  pages={132--164},
  year={2015},
  publisher={Wiley Online Library}
}

@inproceedings{aggarwal2012event,
  title={Event detection in social streams},
  author={Aggarwal, Charu C and Subbian, Karthik},
  booktitle={Proceedings of the 2012 SIAM international conference on data mining},
  pages={624--635},
  year={2012},
  organization={SIAM}
}

@article{hu2017adaptive,
  title={Adaptive online event detection in news streams},
  author={Hu, Linmei and Zhang, Bin and Hou, Lei and Li, Juanzi},
  journal={Knowledge-Based Systems},
  volume={138},
  pages={105--112},
  year={2017},
  publisher={Elsevier}
}

@article{zhou2014event,
  title={Event detection over twitter social media streams},
  author={Zhou, Xiangmin and Chen, Lei},
  journal={The VLDB journal},
  volume={23},
  number={3},
  pages={381--400},
  year={2014},
  publisher={Springer}
}

@inproceedings{xing2016hashtag,
  title={Hashtag-based sub-event discovery using mutually generative lda in twitter},
  author={Xing, Chen and Wang, Yuan and Liu, Jie and Huang, Yalou and Ma, Wei-Ying},
  booktitle={Proceedings of the AAAI conference on artificial intelligence},
  volume={30},
  number={1},
  year={2016}
}

@article{wang2016using,
  title={Using hashtag graph-based topic model to connect semantically-related words without co-occurrence in microblogs},
  author={Wang, Yuan and Liu, Jie and Huang, Yalou and Feng, Xia},
  journal={IEEE Transactions on Knowledge and Data Engineering},
  volume={28},
  number={7},
  pages={1919--1933},
  year={2016},
  publisher={IEEE}
}

@inproceedings{fedoryszak2019real,
  title={Real-time event detection on social data streams},
  author={Fedoryszak, Mateusz and Frederick, Brent and Rajaram, Vijay and Zhong, Changtao},
  booktitle={Proceedings of the 25th ACM SIGKDD international conference on knowledge discovery \& data mining},
  pages={2774--2782},
  year={2019}
}

@article{liu2020story,
  title={Story forest: Extracting events and telling stories from breaking news},
  author={Liu, Bang and Han, Fred X and Niu, Di and Kong, Linglong and Lai, Kunfeng and Xu, Yu},
  journal={ACM Transactions on Knowledge Discovery from Data (TKDD)},
  volume={14},
  number={3},
  pages={1--28},
  year={2020},
  publisher={ACM New York, NY, USA}
}

@inproceedings{liu2018event,
  title={Event detection via gated multilingual attention mechanism},
  author={Liu, Jian and Chen, Yubo and Liu, Kang and Zhao, Jun},
  booktitle={Proceedings of the AAAI conference on artificial intelligence},
  volume={32},
  number={1},
  year={2018}
}

@inproceedings{ren2022evidential,
  title={Evidential temporal-aware graph-based social event detection via dempster-shafer theory},
  author={Ren, Jiaqian and Jiang, Lei and Peng, Hao and Liu, Zhiwei and Wu, Jia and Yu, Philip S},
  booktitle={2022 IEEE International Conference on Web Services (ICWS)},
  pages={331--336},
  year={2022},
  organization={IEEE}
}

@article{ren2023uncertainty,
  title={Uncertainty-guided boundary learning for imbalanced social event detection},
  author={Ren, Jiaqian and Peng, Hao and Jiang, Lei and Liu, Zhiwei and Wu, Jia and Yu, Zhengtao and Yu, Philip S},
  journal={IEEE Transactions on Knowledge and Data Engineering},
  volume={36},
  number={6},
  pages={2701--2715},
  year={2023},
  publisher={IEEE}
}

@article{ren2024toward,
  title={Toward cross-lingual social event detection with hybrid knowledge distillation},
  author={Ren, Jiaqian and Peng, Hao and Jiang, Lei and Hao, Zhifeng and Wu, Jia and Gao, Shengxiang and Yu, Zhengtao and Yang, Qiang},
  journal={ACM Transactions on Knowledge Discovery from Data},
  volume={18},
  number={9},
  pages={1--36},
  year={2024},
  publisher={ACM New York, NY}
}

@article{guo2024unsupervised,
  title={Unsupervised social event detection via hybrid graph contrastive learning and reinforced incremental clustering},
  author={Guo, Yuanyuan and Zang, Zehua and Gao, Hang and Xu, Xiao and Wang, Rui and Liu, Lixiang and Li, Jiangmeng},
  journal={Knowledge-Based Systems},
  volume={284},
  pages={111225},
  year={2024},
  publisher={Elsevier}
}

@article{zhao2017real,
  title={Real-time multimedia social event detection in microblog},
  author={Zhao, Sicheng and Gao, Yue and Ding, Guiguang and Chua, Tat-Seng},
  journal={IEEE transactions on cybernetics},
  volume={48},
  number={11},
  pages={3218--3231},
  year={2017},
  publisher={IEEE}
}

@article{peng2025adaptive,
  title={Adaptive and robust DBSCAN with multi-agent reinforcement learning},
  author={Peng, Hao and Huang, Xiang and Sun, Shuo and Zhang, Ruitong and Wang, Xizhao},
  journal={IEEE Transactions on Pattern Analysis \& Machine Intelligence},
  year={2026},
  doi={10.1109/TPAMI.2025.3648017},
  url = {https://doi.ieeecomputersociety.org/10.1109/TPAMI.2025.3648017}
}
\balance

\appendix

\section{PROMPT OF \framework{}}
\label{AppdxPrompt}
In this appendix, we provide the complete prompts used in \framework{}. 

\vspace{-2mm}
\subsection{Prompt for Evaluation-LLM}
\label{AppdxPromptE}
This prompt is designed to guide the LLM in extracting structured event information, e.g., event name and keywords, from messages. 
The prompt contains role setting, task instruction, and output constraint, as shown in box ~\hyperref[PromptE]{$\mathrm{Prompt}_E$}.

\vspace{-2mm}
\subsection{Prompt for Detection-LLM}
\label{AppdxPromptD}
This prompt is designed to decide whether a message belongs to an existing event or a new event. 
The prompt contains input description, role setting, task instruction, output constraint, and guidance from the knowledge base, as shown in box ~\hyperref[PromptD]{$\mathrm{Prompt}_D$}.
When asking LLMs, the location of \{knowledge\} will be replaced with relevant events retrieved from the knowledge base.

\section{GLOSSARY OF NOTATIONS}
\label{AppdxGN}
 We summarize the notation used in this paper, along with their corresponding description, in Table~\ref{table:Notations}.

 \section{STRUCTURAL ENTROPY}
 \label{AppdxSE}
 Structural entropy~\cite{li2016structural} is a measure of the uncertainty of the graph structure. 
 It represents the minimum number of bits required to encode a reachable vertex during a single-step random walk on the graph. 
 Structural information theory utilizes the encoding tree $\mathcal{T}$ to measure the structure of the graph $G(V, E)$, which is as follows: 
 
 \begin{tcolorbox}[float=b, colback=gray!2, colframe=gray!20!gray, title=\textbf{Prompt$_E$}, before upper=\parindent2em\relax]
\label{PromptE}
\vspace{-1mm}
``You are an event analysis assistant. Your task is to infer the event names discussed in the provided social media comments and extract keywords related to the event. 

First, carefully read all the COMMENTS and understand the core content they discuss.

Second, summarize a concise and accurate EVENT NAME based on the COMMENTS. 

Third, extract no more than 10 KEYWORDS related to the event from the comments, which should cover the core theme, characters, location, time, or other vital information about the event. Each KEYWORD must be a single word that appears in the COMMENTS. 

Answer in JSON format, including EVENT-NAME (str) and KEYWORDS (list) attributes. Other than that, the answer must not include any other information.''
\vspace{-2mm}
\end{tcolorbox}

\begin{tcolorbox}[float=b, colback=gray!2, colframe=gray!20!gray, title=\textbf{Prompt$_D$}, before upper=\parindent2em\relax]
\label{PromptD}
\vspace{-1mm}
``The knowledge base contains EVENTs and corresponding KEYWORDs. 

You are a social media comment classifier determining which one EVENT the INPUT belongs to in the knowledge base. 

Answer in JSON format, including INPUT and EVENT attributes. Other than that, the answer must not include any other information. 

When all knowledge base content is irrelevant to the INPUT, your EVENT answer must be `Others'. 

When the knowledge base is empty, your EVENT answer must be `Others'. 

Answers don't need to consider chat history. 

Here is the knowledge base: 

    \{knowledge\}
    
The above is the knowledge base.''
\vspace{-2mm}
\end{tcolorbox}

 \begin{enumerate} [left=0pt]
    \item Each node $\alpha \in \mathcal{T}$ corresponds to a partitioning of graph nodes $T_\alpha \subseteq V$. 
    Significantly, the root node $\lambda$ of $\mathcal{T}$ is associated with the entire set of graph nodes, $T_\lambda = V$. And for any leaf node $\gamma$ of $\mathcal{T}$, $\mathcal{T}_\gamma$ contains exactly one graph node from $V$.
    \item For any non-leaf node $\alpha$ in $\mathcal{T}$, let its children be denoted as $\beta_1, ..., \beta_{N_{\alpha}}$, where $N_{\alpha}$ is the number of children of $\alpha$. Then, $(T_{\beta_1}, ..., T_{\beta_{N_{\alpha}}})$ form a partitioning of $T_\alpha$.
 \end{enumerate}
 The structural entropy is defined under the graph $G$ and the encoding tree $T$, as follows:
 \begin{equation}
 H^\mathcal{T}(G) = -\sum_{\alpha \in \mathcal{T},\ \alpha \ne \lambda}\frac{g_\alpha}{vol(G)}\log_2\frac{vol(\alpha)}{vol(\alpha^-)},
 \end{equation}
 where $\alpha^-$ is the parent node of non-root node $\alpha$ in $\mathcal{T}$, the cut $g_\alpha$ is the weight sum of edges with exactly one endpoint in $T_\alpha$, and the volume $vol(\alpha)$, $vol(\alpha^-)$, and $vol(\lambda)$ denote the degrees sum of graph nodes within $T_\alpha$, $T_\alpha^-$, and $T_\lambda$, respectively. 

 The encoding tree corresponding to the minimum structural entropy is regarded as the optimal encoding tree $\mathcal{T}^*$, which captures the essential structure of the graph. 
 Without requiring supervision or a predefined number of clusters, the two-dimensional structural entropy minimization algorithm obtains such an optimal encoding tree by an MERGE operator, described as follows:
 \begin{definition}[\text{MERGE} Operator~\cite{li2016structural}]
 Given an encoding tree $\mathcal{T}$ and two of its non-root nodes $\alpha_{i}$ and $\alpha_{j}$, the operation $\text{MERGE}(\alpha_{i}, \alpha_{j})$ removes $\alpha_{i}$ and $\alpha_{j}$ from $\mathcal{T}$ and introduces a new node $\alpha_{n}$ into $\mathcal{T}$. 
 The children of $\alpha_{n}$ are the union of the children of $\alpha_{i}$ and $\alpha_{j}$, and the parent of $\alpha_{n}$ is the root node.
 \end{definition}

 Executing the $\text{MERGE}$ operator will change the structural entropy of $\mathcal{T}$. 
 After initializing the encoding tree, while keeping the tree height no more than 2, the $\text{MERGE}$ operator is repeatedly applied to any two nodes that can largest decrease the structural entropy, until the structural entropy reaches the minimum possible value. 
 This results in the optimal encoding tree, which corresponds to the optimal partitioning of the graph.

 \begin{table}[t]
    \centering
    \caption{Glossary of Notations.}
    \label{table:Notations}
    \setlength{\tabcolsep}{2pt}
    \begin{tabular}{l|l}
    \toprule
   \textbf{Notation} & \textbf{Description} \\
    \midrule
    $\mathcal{S}$ & Social stream, a temporal sequence of message blocks \\
    $\text{M}_t$ & The $t$-th message block in $\mathcal{S}$ \\
    $m_i$ & A social message in block $\text{M}_t$ \\
    $\mathcal{A}_k$ & The $k$-th anchor containing similar messages\\
    $\tau$ & Similarity threshold in each anchor \\
    $\lambda$ & The trade-off parameter of sampling \\
    $p$ & The number of key messages selected from an anchor \\
    $a_k$ & The aggregated key message of anchor $\mathcal{A}_k$ \\
    $y_{m_i}$, $y_{a_k}$ & Event label assigned to $m_i$ and $a_k$ \\
    \midrule
    $\mathcal{KB}^{(t)}$ & Knowledge base at day $t$ used for $\text{M}_t$ \\
    $e$, $p_e$ & Event and its corresponding representation in $\mathcal{KB}$ \\
    $\mathrm{Name}_e$ & Name of event $e$ in $\mathcal{KB}$ \\
    $\mathrm{Keywords}_e$ & Keyword set describing event $e$ in $\mathcal{KB}$ \\
    $\mathrm{Enc}(\cdot)$ & Embedding model for messages and events \\
    $r(e \mid a_k)$ & Relevance score between event $e$ and $a_k$ \\
    $\gamma$ & Similarity threshold in RAG retrieval \\
    $\mathcal{N}_{E(a_i)}$ & Top-$q$ retrieved candidate events for $a_i$ \\
    $\mathcal{B}_e$ & \makecell[l]{Message buffer for event $e$ in knowledge base \\maintenance} \\
    $\theta$ & Threshold of buffered messages for $\mathcal{KB}^{(t)}$ update \\
    \midrule
    $G_t$ & Event graph at day $t$, with nodes, edges, and weights \\
    $H^\mathcal{T}(G_t;\pi_t)$ & Structural entropy of graph $G_t$ with partitioning $\pi_t$ \\
    $\pi^*_t$ & The optimal partitioning of $G_t$ \\
    \bottomrule
 \end{tabular}
 \end{table}

 \section{DATASETS}
 \label{AppdxDataset}
 Following the data processing procedure in~\cite{cao2024hierarchical}, we divide the datasets into daily blocks. 
 The time spans of the Event2012 and Event2018 datasets are 21 days and 16 days, respectively. 
 Dividing into blocks ensures that event detection from the previous day does not use data from the following day, making it more suitable for open-world scenarios. 
 Detailed statistics of the two datasets are shown in Tables~1 and~2.
 
 \begin{table}[h]
 \centering
 \renewcommand{\arraystretch}{0.9}
 \setlength{\tabcolsep}{3.8pt}
 \caption{Detailed statistics of message blocks in Event2012.}
 \label{tab:events2012}
 \vspace{-2mm}
 \begin{tabular}{c|ccccccc}
 \hline
 Blocks & M$_1$ & M$_2$ & M$_3$ & M$_4$ & M$_5$ & M$_6$ & M$_7$ \\
 \hline
 \# Messages & 8,722 & 1,491 & 1,835 & 2,010 & 1,834 & 1,276 & 5278 \\
 \# Events & 41 & 30 & 33 & 38 & 30 & 44 & 57 \\
 \hline
 \hline
 Blocks & M$_8$ & M$_9$ & M$_{10}$ & M$_{11}$ & M$_{12}$ & M$_{13}$ & M$_{14}$ \\
 \hline
 \# Messages & 1,560 & 1,363 & 1,096 & 1,232 & 3,237 & 1,972 & 2,956 \\
 \# Events & 53 & 38 & 33 & 30 & 42 & 40 & 43 \\
 \hline
 \hline
 Blocks & M$_{15}$ & M$_{16}$ & M$_{17}$ & M$_{18}$ & M$_{19}$ & M$_{20}$ & M$_{21}$ \\
 \hline
 \# Messages & 2,549 & 910 & 2,676 & 1,887 & 1,399 & 893 & 2,410 \\
 \# Events & 42 & 27 & 35 & 32 & 28 & 34 & 32 \\
 \hline
 \end{tabular}
 \vspace{-2mm}
 \end{table}

 \begin{table}[h]
 \centering
 \renewcommand{\arraystretch}{0.9}
 \setlength{\tabcolsep}{2.5pt}
 \caption{Detailed statistics of message blocks in Event2018.}
 \label{tab:events2018}
 \vspace{-2mm}
 \begin{tabular}{c|cccccccc}
 \hline
 Blocks & M$_1$ & M$_2$ & M$_3$ & M$_4$ & M$_5$ & M$_6$ & M$_7$ & M$_8$\\
 \hline
 \# Messages & 5,356 & 3,186 & 2,644 & 3,179 & 2,662 & 4,200 & 3,454 & 2,257 \\
 \# Events & 22 & 19 & 15 & 19 & 27 & 26 & 23 & 25 \\
 \hline
 \hline
 Blocks & M$_9$ & M$_{10}$ & M$_{11}$ & M$_{12}$ & M$_{13}$ & M$_{14}$ & M$_{15}$ & M$_{16}$ \\
 \hline
 \# Messages & 3,669 & 2,385 & 2,802 & 2,927 & 4,884 & 3,065 & 2,411 & 1,107 \\
 \# Events & 31 & 32 & 31 & 29 & 28 & 26 & 25 & 14 \\
 \hline
 \end{tabular}
 \vspace{-2mm}
 \end{table}

 \section{IMPLEMENTATION DETAILS}
 \label{AppdxID}
 \framework{} is implemented using the RAGFlow framework, and all experiments are conducted on a server with two NVIDIA RTX 6000 Ada Generation (48GB) GPUs. 
 For the KMS module, we use the ``all-MiniLM-L6-v2'' model (384 dimensions) for embedding the Event2012 dataset and the ``distiluse-base-multilingual-cased-v1'' model (512 dimensions) for the Event2018 dataset. 
 We set the anchor similarity threshold to $\tau = 0.4$ for Event2012 and $\tau = 0.3$ for Event2018, the sampling trade-off parameter to $\lambda = 0.7$, and the number of sampled key messages to $p = 3$. 
 For the event detection module, the RAG retrieval similarity threshold is set to $\gamma = 0$, with a maximum of $q = 8$ candidate events, and the buffering message threshold is set to $\theta = 10$.

\begin{figure*}[t]
\centering
\includegraphics[width=1\linewidth,trim={0cm 0 0cm 0}, clip]{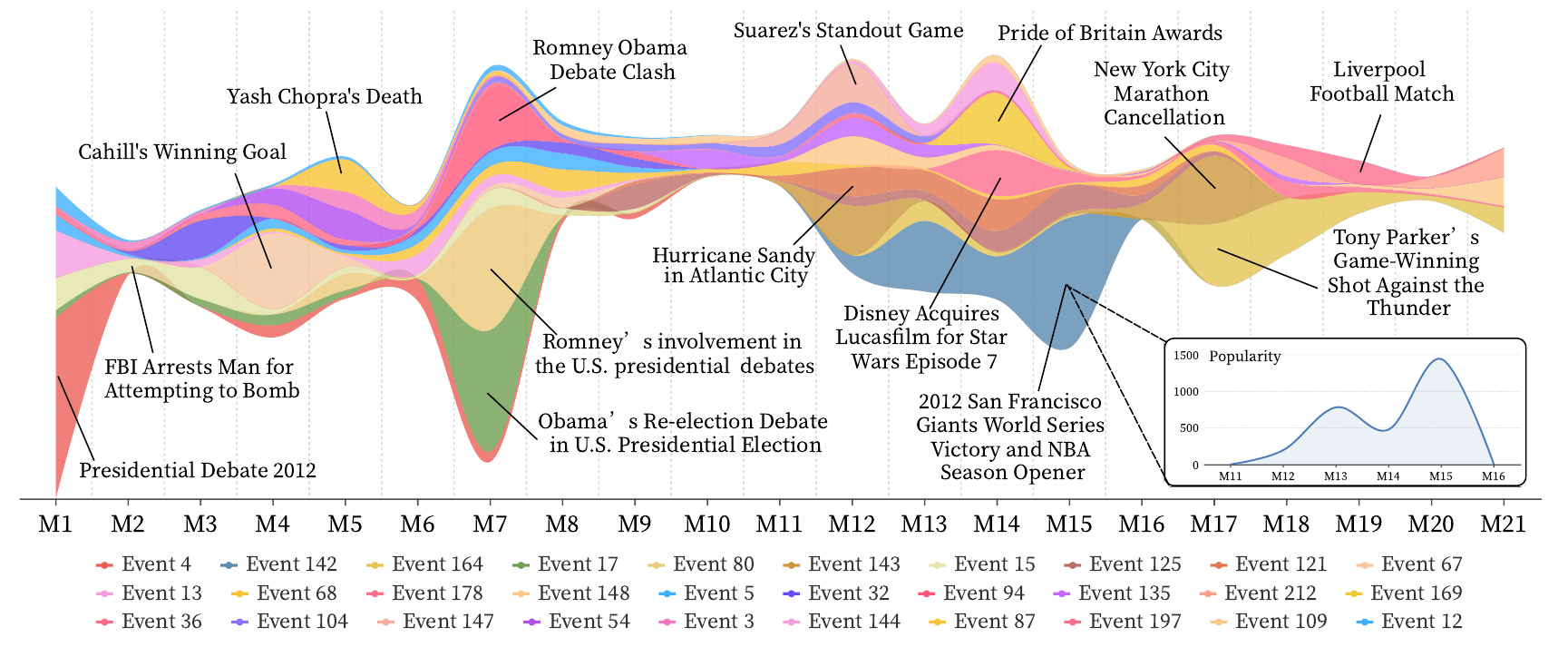}
\vspace{-8mm}
\caption{The visualization of events and their evolution detected by \framework{} on the dataset Event2012.}
\label{fig:visualization2012}
\vspace{-2pt}
\end{figure*}

\begin{figure*}[t]
\centering
\vspace{-2mm}
\includegraphics[width=1\linewidth,trim={0cm 0 0cm 0}, clip]{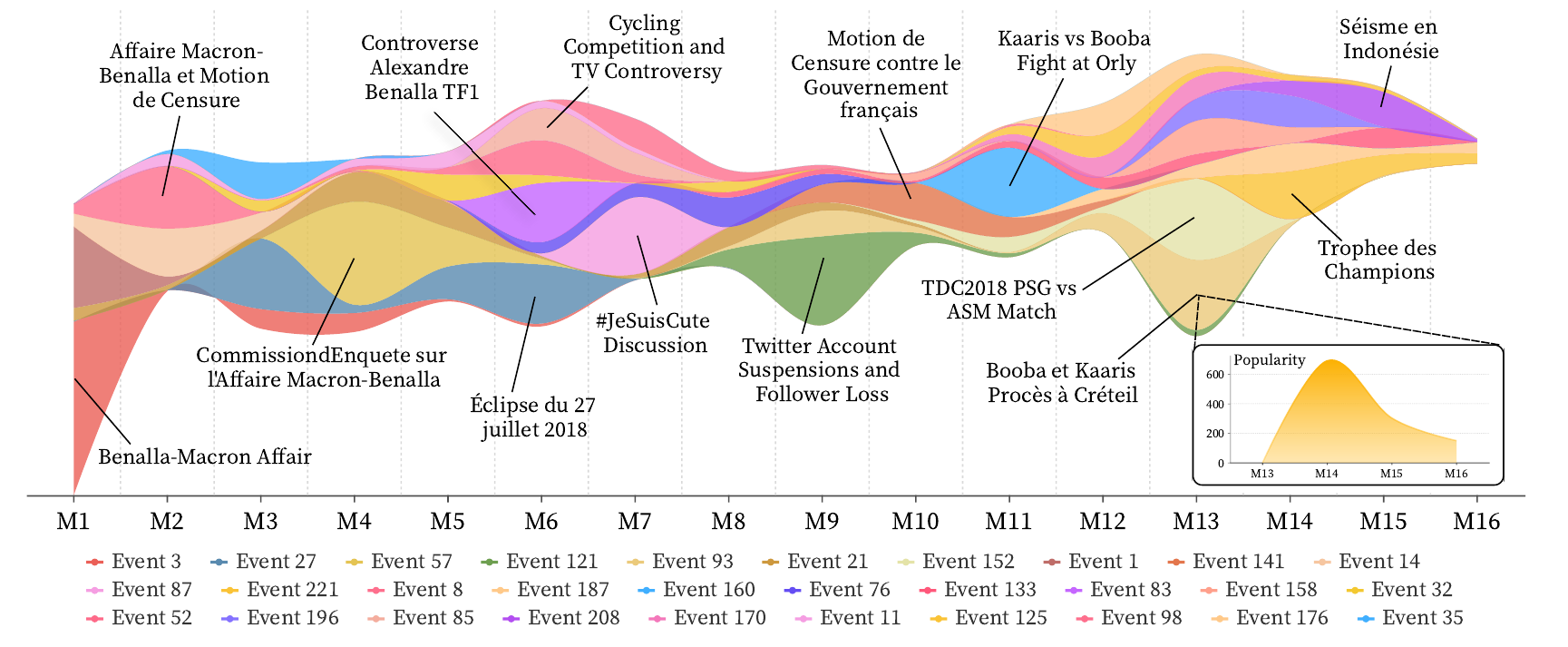}
\vspace{-8mm}
\caption{The visualization of events and their evolution detected by \framework{} on the dataset Event2018.}
\label{fig:visualization2018}
\vspace{-2pt}
\end{figure*}

 \section{BASELINES}
 \label{AppdxBaselines}
 For the SED task, we compare \framework{} with two GNN-based methods, three PLM-based methods, and one structural entropy-based method.
 \textbf{KPGNN} is a knowledge-preserving incremental learning framework for supervised SED using a heterogeneous graph network.
 \textbf{QSGNN} is a framework for self-supervised SED that fine-tunes unlabeled data using pseudo-labels.
 \textbf{SBERT} is a transformer-based model that extends BERT to generate semantically meaningful sentence embeddings.
 We first learn message embeddings using SBERT and then apply K-means clustering on the embeddings to acquire events, following~\cite{cao2024hierarchical}.
 \textbf{CP-Tuning} is an end-to-end contrastive prompt tuning framework for PLMs.
 We use it as~\cite{yu2025promptsed}.
 \textbf{PromptSED} is an evolving topic-enhanced prompt learning framework for SED in an incremental social stream, which does not require additional training or manual labeling.
 \textbf{HISEvent} customizes a hierarchical structure entropy minimization algorithm for unsupervised SED.
 In the main results, we set the hyperparameter $n$ to 200 to mitigate occasional deadlock issues.

 For the SEE task, we compare \framework{} with three non-dynamic topic modeling methods and two dynamic topic modeling methods. 
 All these baselines can extract keywords for events, among which dynamic methods can track the evolution of events.
 \textbf{ProdLDA} introduces autoencoding variational inference for non-dynamic topic modeling.
 \textbf{DecTM} is a causal inference framework to explain and overcome the issues of topic modeling on long-tailed corpora, which is a non-dynamic method.
 \textbf{TSCTM} employs a contrastive learning method to overcome the data sparsity issue in short text non-dynamic topic modeling.
 \textbf{Bertopic} is a dynamic topic modeling method that extracts coherent topic representations through a class-based variation of TF-IDF.
 \textbf{CFDTM} is a chain-free dynamic method for topic modeling using an evolution-tracking contrastive learning.
 
    

    

    


 \section{VISUALIZATION}
 \label{Appdxvi}
 We present the visualizations of the top 30 most discussed social events and their temporal evolution detected by \framework{} on the Event2012 and Event2018 datasets, as shown in Figure~\ref{fig:visualization2012} and Figure~\ref{fig:visualization2018}, respectively.
 In the figures, different colors correspond to different events. The vertical width of each event reflects the discussion intensity, and the “flowing” shape along the horizontal axis illustrates the temporal dynamics of popularity.
 The results show that \framework{} successfully detects a wide range of significant events on both datasets, such as the U.S. presidential election and related debates. 
 More importantly, \framework{} not only captures the initial emergence of these events but also tracks their subsequent developments, resembling the rise and fall of the event streams in the figures. 
 This ability to provide explicit event evolution information distinguishes \framework{} from prior SED methods, which typically fail to model the evolution of detected events.

\end{document}